\documentclass[amsmath,amssymb,aps,prl,reprint,superscriptaddress]{revtex4-1}
\usepackage{graphicx}% Include figure files
\usepackage{xcolor}
\usepackage[colorlinks=true,urlcolor=blue,citecolor=blue,linkcolor=blue]{hyperref}
\usepackage{physics}
\usepackage{siunitx}
\usepackage{xspace}
\usepackage{bm}

\newcommand{\vect}[1]{\vb{#1}}
\newcommand{\transpose}{\mathsf{T}}

\newcommand{\bk}{\vect{k}}
\newcommand{\bq}{\vect{q}}

\newcommand{\bG}{\vect{G}}
\newcommand{\bGp}{\vect{G}'}
\newcommand{\bK}{\vect{K}}
\newcommand{\bKp}{\vect{K}'}
\newcommand{\br}{\vect{r}}

\newcommand{\moire}{moir\'e\xspace}
\newcommand{\TBKT}{T_\text{BKT}}

\usepackage[normalem]{ulem}

\definecolor{TTH-color}{named}{green}
\definecolor{AJ-color}{named}{magenta}
\definecolor{TP-color}{rgb}{0.97,0.57,0.11}
\definecolor{PT-color}{RGB}{128,0,128}

\definecolor{TTH-color2}{named}{green}
\definecolor{AJ-color2}{named}{magenta}
\definecolor{TP-color2}{rgb}{0.87,0.47,0.01}
\definecolor{PT-color2}{RGB}{128,0,128}

\begin{document}

%\preprint{}

\title{Superfluid weight and Berezinskii-Kosterlitz-Thouless transition temperature of twisted bilayer graphene}



%\thanks{}

%\author{A. Julku$^1$}
%\author{T. J. Peltonen$^2$}
%\author{L. Liang$^{1,3}$}
%\author{T. T. Heikkil\"a$^{2}$}
%\email{tero.t.heikkila@jyu.fi}
%\author{P. T\"orm\"a$^{1}$}
%\email{paivi.torma@aalto.fi}
%\affiliation{$^1$Department of Applied Physics, Aalto University, P.O.Box 15100, 00076 Aalto, Finland}
%\affiliation{$^2$\PT{JK, Finland}}
%\affiliation{$^3$Computational Physics Laboratory, Physics Unit, Faculty of Engineering and Natural Sciences, Tampere University, P.O. Box 692, FI-33014 Tampere, Finland}

\newcommand{\affiliationAalto}{Department of Applied Physics, Aalto University, P.O.Box 15100, 00076 Aalto, Finland}
\newcommand{\affiliationJYU}{Department of Physics and Nanoscience Center, University of Jyv\"askyl\"a, P.O. Box 35 (YFL), FI-40014 University of Jyv\"askyl\"a, Finland}
\newcommand{\affiliationTampere}{Computational Physics Laboratory, Physics Unit, Faculty of Engineering and Natural Sciences, Tampere University, P.O. Box 692, FI-33014 Tampere, Finland}

\author{A. Julku}
\affiliation{\affiliationAalto}
\author{T. J. Peltonen}
\affiliation{\affiliationJYU}
\author{L. Liang}
\affiliation{\affiliationAalto}
\affiliation{\affiliationTampere}
\author{T. T. Heikkil\"a}
\email{tero.t.heikkila@jyu.fi}
\affiliation{\affiliationJYU}
\author{P. T\"orm\"a}
\email{paivi.torma@aalto.fi}
\affiliation{\affiliationAalto}


\date{\today}

\begin{abstract}
We study superconductivity of twisted bilayer graphene with local and non-local attractive interactions. We obtain the superfluid weight and Berezinskii-Kosterlitz-Thouless (BKT) transition temperature for microscopic tight-binding and low-energy continuum models. We predict qualitative differences between local and non-local interaction schemes which could be distinguished experimentally. In the flat band limit where the pair potential exceeds the band width we show that the superfluid weight and BKT temperature are determined by multiband processes and quantum geometry of the band.
\end{abstract}

%\keywords{}

\maketitle


%\textit{Introduction}
Recent experimental discoveries of superconductivity in bilayer graphene twisted close to a ``magic angle" $\theta^*$~\cite{cao:2018:1,yankowitz:2019,lu:2019} call for a reconsideration of traditional theories of superconductivity~\cite{kopnin:2011,ojajarvi:2018}, in particular because the superconductivity occurs in a regime where the non-interacting electronic states form an asymptotically flat (dispersionless) band~\cite{li:2010,bistritzer:2010,shallcross:2010,suarezmorell:2010,brihuega:2012,lopesdossantos:2012,tramblydelaissardiere:2012,shallcross:2013,uchida:2014,fang:2016,weckbecker:2016,nam:2017}. As the system is two-dimensional, the transition to superconductivity is bound to occur at the Berezinskii-Kosterlitz-Thouless (BKT) temperature $\TBKT$~\cite{berezinskii:1972,kosterlitz:1972,kosterlitz:1973} which can be determined from $k_B\TBKT=\frac{\pi}{8} \sqrt{\det[D^s(\TBKT)]}$~\cite{nelson:1977,cao:2014}. Here $D^s$ is the superfluid weight that yields the size of the supercurrent for a given phase gradient of the order parameter. In conventional theory of superconductivity \cite{tinkham2004introduction}, $D^s$ is proportional to the group velocity of electronic bands around the Fermi level. Thus $D^s = 0$ for a flat band, and superconductivity in twisted bilayer graphene (TBG) appears puzzling. One might argue it to be due to the bands not being perfectly flat; however, we show here that a more likely explanation goes beyond the conventional theory. Here we calculate  \textit{$\TBKT$  for TBG as function of the superconducting order parameter and filling}. We use two models of TBG including both the flat and a number of dispersive bands
and show that superconductivity in the flat band regime has essentially a \textit{quantum geometric origin}. 

Recently, it was found that $D^s$ has, in addition to the conventional contribution proportional to group velocity, a geometric contribution arising from multiband processes~\cite{peotta:2015,julku:2016,Huber:2016,liang:2017,Torma:2018}. In a flat band limit the geometric contribution dominates and is bounded from below by the band Berry curvature~\cite{liang:2017} and Chern number~\cite{peotta:2015}. Here we show that the geometric contribution dominates $D^s$ and $T_{\rm BKT}$ in the flat band regime of TBG. Importantly, we show that \emph{including only the few flat bands is not sufficient} but one needs also a number of dispersive bands to correctly predict the geometric contribution. Therefore, approximate models of TBG such as those with only flat bands, as used for deriving upper \cite{hazra:2018} and lower \cite{xie:2019} bounds of the superfluid weight and in many other works \cite{guo:2018,lian:2018,chen:2018,xu:2018,liu:2018,tang:2019,roy:2018,sherkunov:2018,lin:2018,classen:2019,ray:2019,dodaro:2018,kang:2019}, may not be suited for quantitative predictions of %the transition temperature and supercurrent
TBG superconductivity. Moreover, we predict that, in the flat-band regime, local ($s$-wave) and non-local interactions yield distinct behavior, namely an \emph{anisotropic superfluid weight} in the latter case. We propose a four-terminal radio frequency spectroscopy \textit{experiment} that can detect the possible anisotropy and thus \textit{distinguish between the two pairing mechanisms}. 

\begin{figure}
\includegraphics[width=1.0\columnwidth]{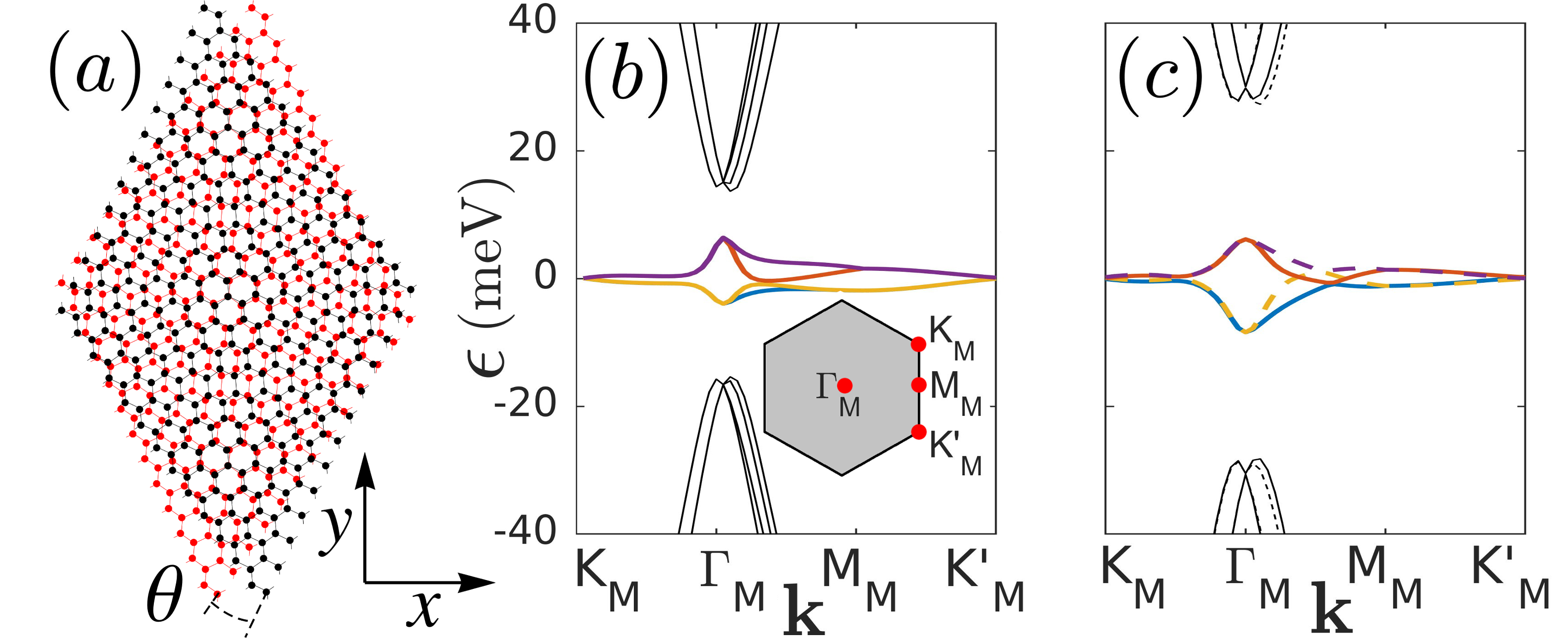}
\caption{(a) The \moire superlattice of TBG depicted with a twist angle $\theta$ and the choice of the $x$ and $y$-axes. (b)--(c) Single-particle energy band structures of the RM and DP methods, respectively, plotted within the \moire Brillouin zone along the path connecting the high symmetry points shown in the inset of (b). In the DP model (c) the bands coming from the valley $\bK$ ($\bKp$) are drawn as solid (dashed) lines.}
\label{fig:1}
\end{figure}

%One of the 
An outstanding problem in describing the TBG physics theoretically~\cite{liu:2018,huang:2019,ray:2019,kozii:2019,liu:2019,guo:2018,kennes:2018,choi:2018,peltonen:2018,wu:2018:2,lian:2018,wu:2019,wu:2019:2,wu:2019:3,sherkunov:2018,su:2018,xu:2018,roy:2018,po:2018,isobe:2018,laksono:2018,fidrysiak:2018,tang:2019,venderbos:2018,dodaro:2018,lin:2018,gonzalez:2019,classen:2019,chen:2018} is the fact that the unit cell of the \moire superlattice with twist angles close to $\theta^*$ contains a large amount of carbon atoms [Fig. \ref{fig:1}(a)], and therefore TBG theory should take a stand on how to describe the interlayer couplings within this unit cell. Here we use and compare with each other two of the previously used approximation procedures: (1) the \textit{renormalized \moire} (RM) approach \cite{gonzalez-arraga:2017,su:2018}, where we scale some coupling energies by a suitable scaling factor to find the flat bands at a higher $\theta$, resulting into a smaller unit cell, and (2) the \textit{Dirac point approximation} (DP) \cite{peltonen:2018,lopesdossantos:2007,lopesdossantos:2012},
where we make a low energy approximation near the graphene Dirac points by linearizing the intralayer Hamiltonians and using a cut-off in the superlattice Fourier space. Both of these approaches go beyond those often used in TBG literature, either based on a single-parameter coupling model, or a vastly reduced four-band model \cite{yuan:2018,kang:2018,koshino:2018,hazra:2018,xie:2019}.

\textit{Theoretical models} 
In the renormalized \moire lattice method (RM), we deploy the Fermi-Hubbard Hamiltonian as \cite{supplement} $H = H_{\textrm{kin}} -\mu N + H_{\textrm{int}}$, where $H_{\textrm{kin}} = \sum_{i\alpha j\beta\sigma}t_{i\alpha j\beta}c^\dag_{i\alpha\sigma}c_{j\beta\sigma}$ is the kinetic term, $N$ is the total particle number operator and $H_{\textrm{int}}$ is the effective attractive interaction described below. Here $c_{i\alpha\sigma}$ annihilates a fermion in the $\alpha$th lattice site of the $i$th \moire superlattice unit cell with spin $\sigma \in \{\uparrow,\downarrow \}$, $\mu$ is the chemical potential and the hopping $t_{i\alpha j\beta}$ includes both the intra- and interlayer terms.

Since the type of the interaction is not currently known, we consider two different singlet pairing potentials, namely the local pairing $H_{\textrm{int}} = J\sum_{i\alpha}c^\dag_{i\alpha\uparrow}c^\dag_{i\alpha\downarrow}c_{i\alpha\downarrow}c_{i\alpha\uparrow} \equiv H_{\textrm{loc}}$ and the nearest-neighbour (NN) pairing $H_{\textrm{int}} = \frac{J}{2} \sum_{\langle i\alpha j\beta \rangle} h^\dag_{i\alpha j\beta} h_{i\alpha j\beta} \equiv H_{\textrm{RVB}}$, where $h_{i\alpha j\beta} = (c_{i\alpha \uparrow}c_{j\beta \downarrow} - c_{i\alpha \downarrow}c_{j\beta \uparrow})$ and $J<0$ is the interaction strength. The local interaction has been used to study $s$ wave superconductivity, mediated by electron-phonon interaction, both in graphene \cite{zhao:2006,uchoa:2007,kopnin:2008,hosseini:2015} and in TBG \cite{peltonen:2018,wu:2018:2}. The non-local, called resonance valence bond (RVB) interaction \cite{anderson:1987,lee:2006}, has also been used both in case of monolayer graphene \cite{baskaran:2002,black-schaffer:2007,pathak:2010,black-schaffer:2012} and TBG \cite{su:2018}. We keep only the pairing channels by applying mean-field theory to approximate $H_{\textrm{loc}} \approx \Delta_{i\alpha} c_{i\alpha\uparrow}^\dag c_{i\alpha\downarrow}^\dag + \textrm{H.c.}$ and $H_{\textrm{RVB}} \approx \Delta_{i\alpha j\beta} h_{i\alpha j\beta}^\dag  + \textrm{H.c.}$, where $\Delta_{i\alpha} = J\langle c_{i\alpha\downarrow}c_{i\alpha\uparrow} \rangle$ and $\Delta_{i\alpha j\beta} = \frac{J}{2}\langle h_{i\alpha \beta} \rangle$  are the superfluid order parameters, respectively.

To reduce the number of lattice sites $M$ within a \moire unit cell (around 12 000 for twist angle $\theta \sim \SI{1}{\degree}$), we apply a rescaling trick \cite{su:2018,gonzalez-arraga:2017} under which the Fermi velocity of a monolayer graphene and the \moire periodicity remain invariant but  $\theta$ becomes larger and thus reduces $M$. In our computations we use the rescaling such that $M=676$ and the rescaled angle is $\theta' = \SI{4.41}{\degree}$ \cite{supplement} which reproduces the four narrow bands of the bandwidth of $\SI{10}{meV}$ found experimentally with $\theta \sim \SI{1}{\degree}$ [see Fig. \ref{fig:1}(b)].

In the Dirac point continuum method (DP) we employ the low-energy \cite{lopesdossantos:2012,peltonen:2018,supplement} Dirac point approximation for the two graphene layers as $H_\text{kin}^1 = \sum_{\sigma\rho\bk\vect{G}} c_{\sigma\rho,1}^\dagger(\bk+\bG) \hbar v_F \bm{\sigma}^\rho\vdot(\bk+\bG) c_{\sigma\rho,1}(\bk+\bG)$ and $H_\text{kin}^2 = \sum_{\sigma\rho\bk\vect{G}} c_{\sigma\rho,2}^\dagger(\bk+\bG) \hbar v_F \bm{\sigma}_\theta^\rho\vdot(\bk+\bG) c_{\sigma\rho,2}(\bk+\bG)$ and couple the layers by $H_\text{kin}^\perp = \sum_{\sigma\rho\bk\bG\bGp} c_{\sigma\rho,1}^\dagger(\bk+\bG+\rho\frac{\Delta\bK}{2}) t_\perp^\rho(\bG-\bGp) c_{\sigma\rho,2}(\bk+\bGp-\rho\frac{\Delta\bK}{2}) +\text{H.c.}$ Here $c_{\sigma\rho,l}(\vect{k}) = (c_{\sigma\rho,lA}(\vect{k}),c_{\sigma\rho,lB}(\vect{k}))^\transpose$ in the sublattice space, where $c_{\sigma\rho,ls}(\bk)$ is the annihilation operator for spin $\sigma$, valley $\rho\in\{+,-\}$, layer $l$, sublattice $s$, and wavevector $\bk$, $\bm{\sigma}^\rho=(\rho\sigma_x,\sigma_y)$ is a vector of Pauli matrices in the sublattice space, $\bm{\sigma}_\theta^\rho = R(\theta)\bm{\sigma}^\rho$ is the $\theta$-rotated version of it, $t_\perp^\rho(\bG)$ is the Fourier component \cite{note:DP} of a Slater-Koster \cite{Slater:1954} parametrized interlayer potential (times an exponential factor), $\Delta\bK=R(\theta)\bK-\bK$ is the difference vector from the graphene $\bK$ point to its rotated counterpart, and $v_F$ is the graphene Fermi velocity. The $\bk$ sum is over the the \moire Brillouin zone and the $\bG,\bGp$ sums are over the (truncated) reciprocal superlattice. 

We then write the total Hamiltonian as $H = H_\text{kin}^1 + H_\text{kin}^2 + H_\text{kin}^\perp - \mu N + H_\text{int}$, where $N$ is the total particle number operator. To describe the superconducting state with a local pairing interaction $\lambda$ we use  $H_\text{int} = \lambda\sum_{ls}\int\dd{\vect{r}} \psi_{\uparrow\rho,ls}^\dagger(\vect{r}) \psi_{\downarrow\bar{\rho},ls}^\dagger(\vect{r}) \psi_{\downarrow\bar{\rho},ls}(\vect{r}) \psi_{\uparrow\rho,ls}(\vect{r})$, which is treated in the mean-field level \cite{supplement}. Here $\bar{\rho}$ is the opposite valley of $\rho$ and $\psi_{\sigma\rho,ls}(\vect{r})$ is the continuum electron field operator.

\begin{figure}
\includegraphics[width=0.80\columnwidth]{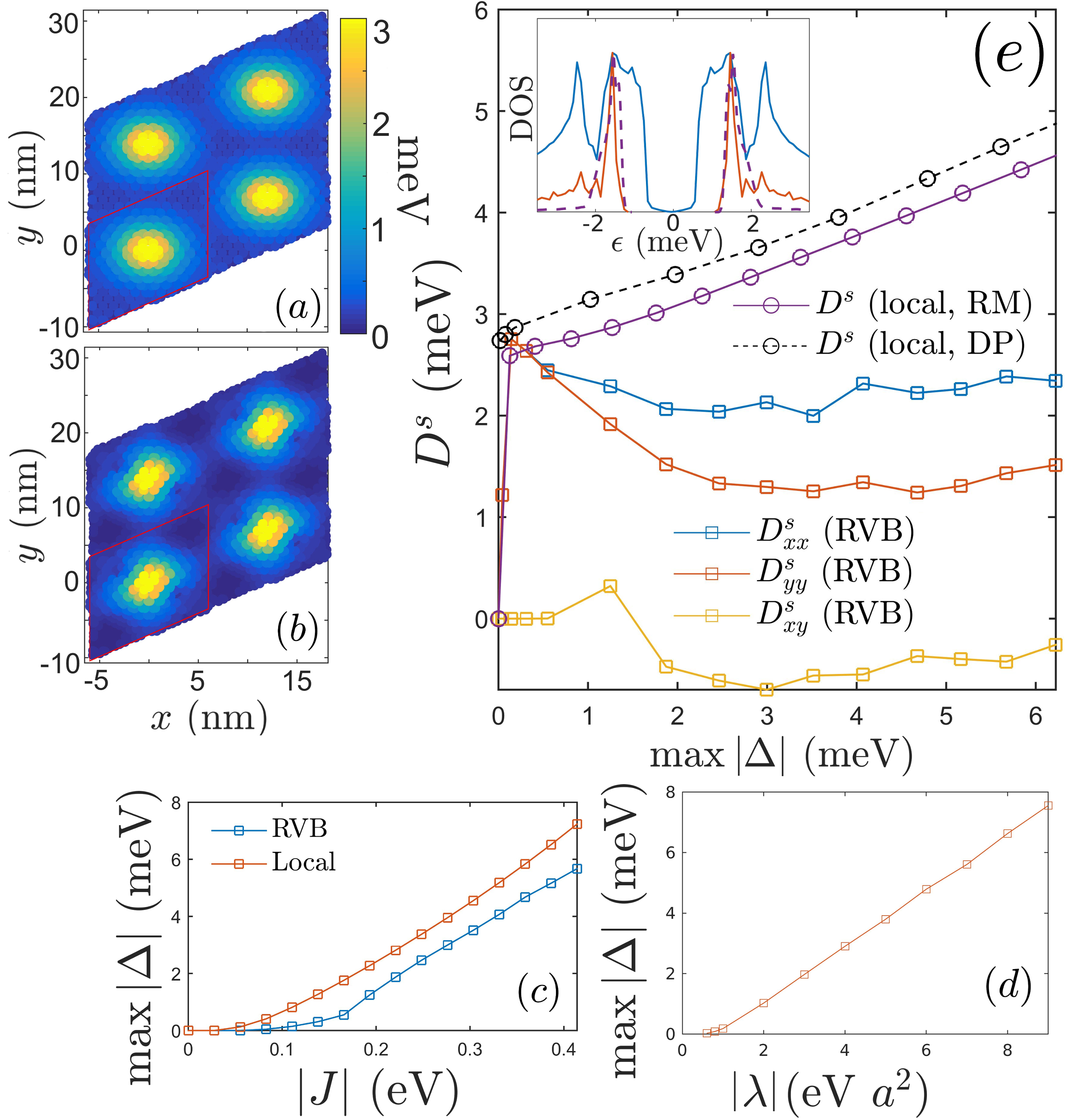}
\caption{(a)--(b) Spatial profiles of the order parameter for local and RVB interaction schemes, respectively, computed with RM. The DP model for the local interaction yields a similar spatial distribution \cite{peltonen:2018}. In case of RVB, $\Delta_{i\alpha j\beta}$ are plotted at $\br_{i\alpha}$. Red parallelograms represent the \moire unit cell. The maximum order parameter in both cases is $\max|\Delta| \approx \SI{3.4}{meV}$. (c)--(d) $\max|\Delta|$ as a function of the interaction strength at $\nu \approx -2$ for the RM and DP methods, respectively. Here $a$ is the graphene lattice constant. (e) Spatial components of $D^s$ as a function of $\max |\Delta|$ at $\nu \approx -2$ for local and RVB pairing. For local interaction $D^s_{xx} = D^s_{yy} = D^s$. Inset of (e) shows the total density of states (DOS) for RVB (blue curve) and local interaction (red) at $\max |\Delta| \approx \SI{3.4}{meV}$ computed with RM. The dashed curve is the DOS for local interaction obtained with DP at $\max |\Delta| \approx \SI{3.5}{meV}$. From the DOS we see the nematic phase being gapless, while the $s$ wave state is gapped. The RM results are evaluated at $T\approx \SI{0.1}{K}$, whereas the DP results at $T=0$.}
\label{fig:2}
\end{figure}

\begin{figure}
\includegraphics[width=1.0\columnwidth]{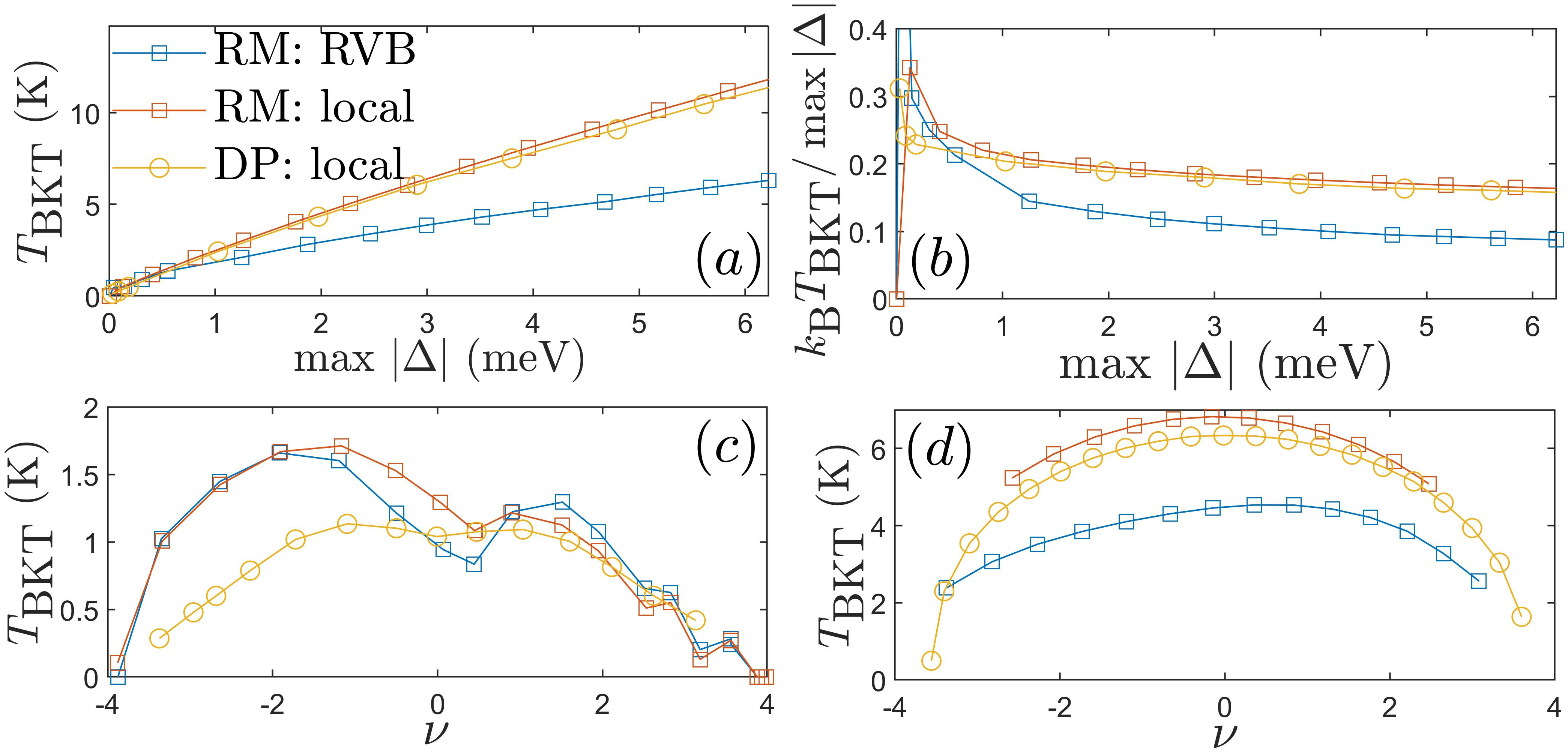}
\caption{(a)--(b) $\TBKT$ and $k_B \TBKT/\max |\Delta(T=0)|$, respectively, as a function of $\max |\Delta(T=0)|$ at $\nu \approx -2$. This result is almost independent of the filling \cite{supplement}. (c)--(d) $\TBKT$ as a function of $\nu$ at $\max|\Delta| \approx \SI{0.4}{meV}$  and $\max|\Delta| \approx \SI{3}{meV}$ at CNP, respectively.}
\label{fig:3}
\end{figure}

\textit{Order parameters, superfluid weight, and pairing symmetry} In experiments \cite{cao:2018:1,yankowitz:2019,lu:2019} superconducting (SC) and correlated insulating states have been observed with the magic angle twist such that insulating states emerge for the flat band fillings $\nu \in \{0, \pm 1, \pm 2, \pm 3 \}$ and SC states surround the insulating states close to $\nu \in \{0,\pm 1, \pm 2\}$ with the SC phase near $\nu = -2$ being observed at temperature as high as $\sim \SI{3}{K}$ \cite{yankowitz:2019,lu:2019}. Here $\nu$ is the electron density per \moire unit cell so that the charge neutrality point (CNP) corresponds to $\nu = 0$ and narrow bands are empty (full) when $\nu = -4$ ($\nu = 4$). 

To determine the superfluid weight $D^s$, we first solve order parameters from the BCS gap equations \cite{supplement}. In Figs. \ref{fig:2}(a)--(b) we show the spatial profiles for the local and RVB interactions computed with the RM method at $\nu \approx -2$. Here $J$ is chosen such that the maximum value of the order parameter is $\max|\Delta| \approx \SI{3.4}{meV}$. From Figs. \ref{fig:2}(c)--(d) we see $\max|\Delta|$ depending almost linearly on the interaction constant, which is typical for generic flat band systems \cite{kopnin:2011,peotta:2015,julku:2016,ojajarvi:2018,peltonen:2018}. From the obtained order parameter values one can compute $D^s$. For easier comparison between the RM and DP models, below we use $\max|\Delta|$ as a ``parameter''.

To obtain $D^s$ we use linear response theory. In the mean-field level \cite{scalapino:1992,scalapino:1993} the zero-frequency, long wavelength limit of the current-current response function $K_{\mu\nu}(\bq,\omega)$ is $D^s$, i.e. ($\mu,\nu\in\{x,y\}$)
\begin{equation}
\label{eq:1}
D^s_{\mu\nu} = \lim_{\bq \rightarrow 0} \big[ \lim_{\omega \rightarrow 0} K_{\mu\nu}(\bq,\omega) \big].
\end{equation}
In Ref. \onlinecite{liang:2017} this was computed for a generic multi-orbital lattice geometry with the local interaction. The details on how $D^s_{\mu\nu}$ is obtained for our different models are discussed in the Supplementary Material (SM) \cite{supplement}.

In Fig. \ref{fig:2}(e) we present $D^s$ as a function of $\max|\Delta|$ at $\nu=-2$ for both local (obtained with RM and DP) and RVB interactions (only RM). Figs. \ref{fig:2}(a)--(b) and \ref{fig:2}(e)  depict a striking distinction between the local and RVB pairing schemes related to the pairing symmetry and the resulting form of $D^s$. The local pairing, yielding an $s$-wave symmetry, conserves the underlying $C_3$-symmetry of the TBG lattice [Fig. \ref{fig:2}(a)] and $D^s$ is isotropic \cite{supplement}, i.e. $D^s_{xx} = D^s_{yy}$ and $D^s_{xy} = D^s_{yx} = 0$. By contrast, the RVB pairing with strong enough interaction breaks the $C_3$-rotational symmetry and yields nematic pairing pattern in real space [Fig. \ref{fig:2}(b)] which leads to an anisotropic response, i.e. $D^s_{xx} \neq D^s_{yy}$ and $D^s_{xy} = D^s_{yx} \neq 0$. The $s$-wave is gapped, whereas the nematic phase has nodal points in the \moire Brillouin zone [see also the inset of Fig. \ref{fig:2}(e)]. The anisotropic $D^s$ results into an anisotropic kinetic inductance of TBG, and it can in principle be accessed via radio frequency impedance spectroscopy \cite{chiodi2011probing} in a Hall-like four-probe setup.

As seen from Fig. \ref{fig:2}(e), $D^s$ for the RVB interaction in the weak-coupling regime is still isotropic. This phase has the mixed $(d+id) + (p+ip)$ symmetry with a full energy gap, whereas the nematic phase of the flat band regime is identified as a mixture of $s$, $p$ and $d$-wave components \cite{su:2018}, with the $d$-wave being the dominant symmetry. Our results for the pairing symmetry are in agreement with Ref. \onlinecite{su:2018} and they differ from the topological $d+id$ symmetry predicted in many TBG studies~\cite{guo:2018,xu:2018,liu:2018,fidrysiak:2018,huang:2019,angeli:2019,wu:2019,laksono:2018,lin:2018,classen:2019} and also from other proposed symmetries which include $s$-wave \cite{peltonen:2018,wu:2018:2,choi:2018,po:2018,wu:2019:3,laksono:2018}, extended $s$-wave \cite{ray:2019,liu:2019,sherkunov:2018,angeli:2019}, $p$-wave \cite{isobe:2018,wu:2019:3}, $p+ip$-wave \cite{roy:2018}, $d$-wave \cite{wu:2018:2,isobe:2018,wu:2019:3}, and $f$-wave \cite{tang:2019,wu:2019:3,lin:2018,classen:2019}. Apart from Ref. \onlinecite{su:2018}, nematic pairing has been predicted only in a few works~\cite{dodaro:2018,wu:2019,wu:2019:2,venderbos:2018,roy:2018}. The microscopic RM method allows to find the nematic pairing, unlike four-band models of Refs. \cite{yuan:2018,kang:2018,koshino:2018}.

\textit{BKT-transition temperature} By computing $D^s$, one can determine $\TBKT$. In Fig. \ref{fig:3}(a) we show $\TBKT$ as a function of  $\max|\Delta|$. We can distinguish two qualitatively different regimes: in the weak-coupling limit the RVB and local interactions yield similar $\TBKT$ whereas for stronger interactions $\TBKT$ depends on the pairing model. Moreover, around $\max|\Delta|\gtrsim \SI{2}{meV}$ the behaviour of the $\TBKT$ curves is almost linear, in accordance with previous studies \cite{peotta:2015,julku:2016} where $D^s$ of a flat band with the local interaction was shown to depend linearly on the pairing strength. In our case the narrow bands are not exactly flat but slightly dispersive and thus their flat band characteristics manifest only when the interaction strength is sufficiently large \cite{peltonen:2018}. Because of this, we call the regime with $\max|\Delta|\gtrsim \SI{2}{meV}$ as the flat band limit. In this regime the DP and RM results are in agreement, whereas for weak interactions the results differ due to different band structures.

The difference of the two interaction schemes is further highlighted in Fig. \ref{fig:3}(b) which presents the ratio $k_B \TBKT/\max |\Delta(T=0)|$. At the flat band limit this ratio approaches a constant whose value depends on the pairing potential. In experiments one can measure $\TBKT$ and in principle also deduce $\Delta$ (from the local density of states) and thus the ratio of these two quantities can be used %could be of relevance 
to characterize the SC pairing observed in experiments.

In Figs. \ref{fig:3}(c)--(d) we present $\TBKT$ as a function of $\nu$. The weak-coupling regime shows a dome-shaped structure of $\TBKT$ which reaches its maxima near the half-fillings of the hole- and electron-doped regimes, similar to experiments \cite{lu:2019}. In the RM model the hole-doped region is much stronger due to higher density of states at negative energies [see Fig. \ref{fig:1}(b)], while the DP model exhibits approximate electron-hole symmetry. Strong asymmetry of RM model is due to the applied rescaling approximation which amplifies the finite but small asymmetry of the unscaled model \cite{supplement}.  In the flat band limit, the shape of the one-particle dispersions are, except for the pronounced  particle-hole  asymmetry of the RM model, completely  dissolved.

\textit{Geometric contribution} One can decompose $D^s$ to conventional, $D^s_{\textrm{conv}}$, and geometric, $D^s_{\textrm{geom}}$, parts, so that $D^s = D^s_{\textrm{conv}} + D^s_{\textrm{geom}}$ \cite{peotta:2015,liang:2017,supplement}. The conventional term depends on the inverse of the effective mass of the Bloch bands and is thus a single-band contribution, whereas  $D^s_{\textrm{geom}}$  is a multiband effect depending on the overlap of the Bloch states and their momentum derivatives  of the form $\braket{\partial_{\bk} n}{m}$, where $\ket{m}$ are the single-particle states of the $m$th Bloch band and $n\neq m$ \cite{liang:2017}, i.e. $D^s_{\textrm{geom}} = 0$ for a single-band system. For a strictly flat band, $D^s_{\textrm{conv}}=0$ so its superconductivity is purely a multiband process characterized by a finite $D^s_{\textrm{geom}}$. This arises an intriguing question related to TBG system: how much the interband terms between dispersive and narrow bands affect $D^s$ via $D^s_{\textrm{geom}}$? 

\begin{figure}
\includegraphics[width=1.0\columnwidth]{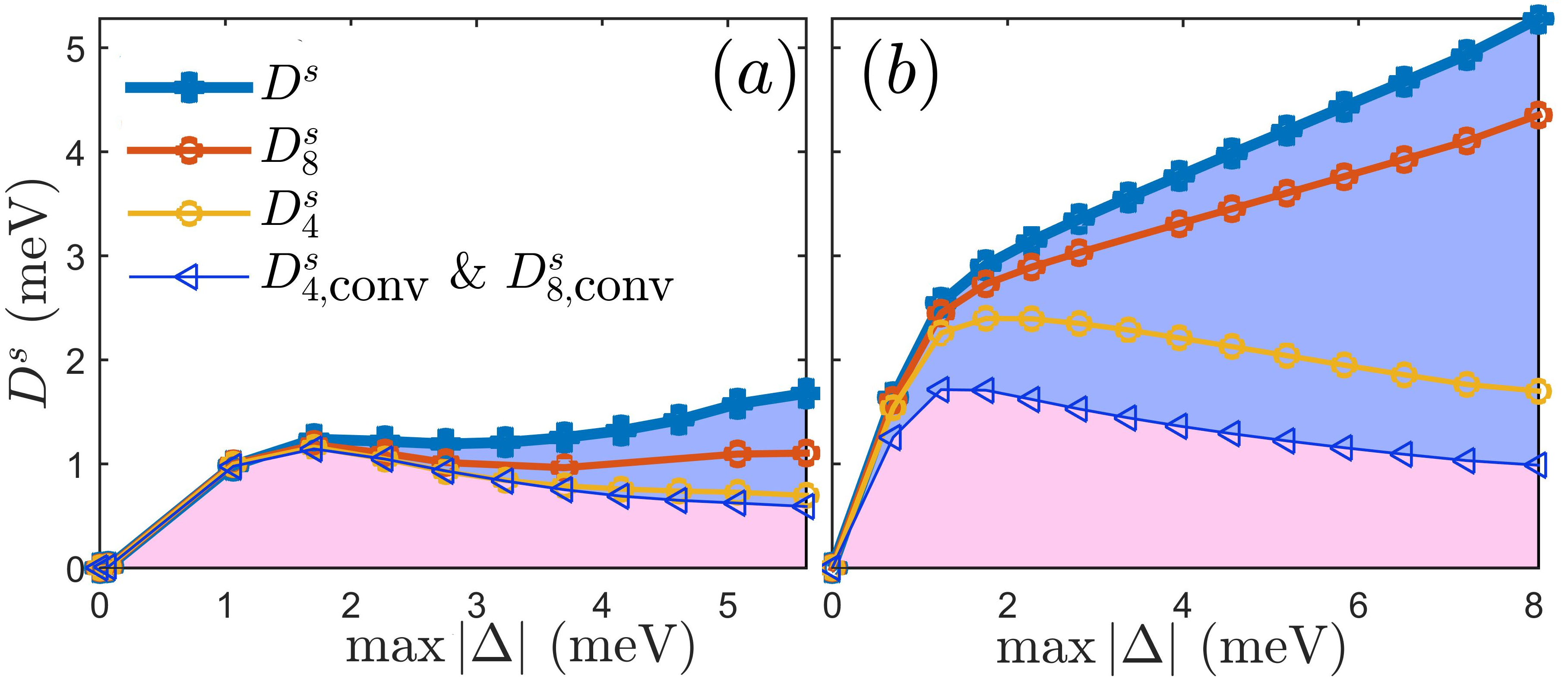}
\caption{Various superfluid components as a function of $\max |\Delta|$ at $\nu \approx -2$ and $T = \SI{1.5}{K}$ for the (a) RVB and (b) local interaction obtained from the RM model. Blue curve is $D^s$ and blue (pink) area depicts $D^s_{\textrm{geom}}$ ($D^s_{\textrm{conv}}$). Results for $D^s$ and  $D^s_{\textrm{conv}}$ computed by considering only $4$ and $8$ Bloch bands are also shown, labeled as $D^s_4$, $D^s_8$ and $D^s_{4/8,\textrm{conv}}$. We have numerically checked that $D^s_{\textrm{conv}} \approx D^s_{4,\textrm{conv}}$.}
\label{fig:4}
\end{figure}

We study this question in Fig. \ref{fig:4} for RVB [Fig. \ref{fig:4}(a)] and local [Fig. \ref{fig:4}(b)] pairing by presenting the total $D^s$ and its components. We further show results obtained by taking into account \cite{supplement} either only the $4$ flat bands or the 8 lowest (4 flat, 4 dispersive) bands  labeled as $D^s_4$ and $D^s_8$, respectively. In both pairing cases the contribution coming from the 4 flat bands only is relatively small for larger interactions. The contribution of 8 bands is larger due to a larger $D^s_{\textrm{geom}}$, which is caused by the interband terms between dispersive and flat bands as the terms between the dispersive bands only are negligible. The slight dispersion of the narrow bands results in a finite $D^s_{\textrm{conv}}$. Note that $D^s_{\textrm{conv}} \approx D^s_{4,\textrm{conv}}$, i.e. $D^s - D^s_{4,\textrm{conv}}$ gives the total $D^s_{\textrm{geom}}$. From Fig. 4 we see that, at $\max |\Delta| \sim \SI{1\dots2}{meV}$ i.e. when the system enters the flat band regime, $D^s_{\textrm{geom}}$ surpasses $D^s_{\textrm{conv}}$ for local pairing and becomes significant in the RVB case. An important implication of Fig. \ref{fig:4} is \textit{the importance of the dispersive bands when computing $D^s$ and the insufficiency of four band models}. Also for a non-interacting system, higher bands have been argued to be necessary, but for different (symmetry) reasons~\cite{po:2019}.

\textit{Discussion}
Our work shows that TBG is characterized by two distinct superconducting regimes. When $\Delta$ is much smaller than the flat band bandwidth, the superfluid weight $D^s$ and the BKT transition temperature $\TBKT$ are well described by conventional theory of superconductivity. On the other hand, in this weak coupling regime the results are somewhat different for the RM and DP models. This is consistent with the low energy dispersion in TBG being very sensitive to the details of the model used \cite{walet:2019}.
In the flat band regime where $\Delta$ is larger than the width of the significant density of states in the flat bands, a major contribution to the superfluid weight $D^s$ originates from the geometric properties of the bands. The geometric contribution $D^s_{\textrm{geom}}$ is proportional to the quantum metric~\cite{peotta:2015} whose importance in physics has been recently emerging
\cite{Bengtsson:2006,Gu:2010,Neupert:2013,Roy:2014,Regnault:2013,peotta:2015,Srivastava:2015,Gao:2015,julku:2016,Huber:2016,Montambaux:2016,liang:2017,Torma:2018,Ozawa:2018,Hughes:2019,Yu:2018,Gianfrate:2019,Weitenberg:2019,Yang:2019}. Moreover, in the flat band regime, both $D^s$ and $\TBKT$ depend sensitively on the pairing mechanism, but not strongly on the employed microscopic model. In particular, for a non-local RVB interaction $D^s$ becomes anisotropic, which could be seen in four-terminal radio frequency spectroscopy experiments to reveal information about the pairing mechanism. 

Within both of our models, at $\theta^*$ the crossover between the two regimes takes place for $\Delta=\SI{1\dots 2}{meV}$, implying $\TBKT \approx \SI{1.5 \dots 3}{K}$. This is also the ballpark of the experimentally accessed critical temperatures \cite{yankowitz:2019,lu:2019}. Thus the geometric contribution of the superfluid weight and the dependence on the pairing mechanism should be relevant for current experiments. An interesting future direction of research is to include other interaction channels than pairing and explore the insulating states observed in TBG \cite{cao:2018:2,yankowitz:2019,lu:2019}. Based on our results, one can anticipate that quantum geometry and multiband processes are important in superconductivity and correlated states of other twisted multilayer materials \cite{Taniguchi:2019_1,Taniguchi:2019_2,Kim:2019,Cao:2019_2,Zhang:2018_1,
Zhang:2018_2,Senthil:2018,MacDonald:2019,Senthil:2019_1,Senthil:2019_2,Dai:2019,Vishwanath:2019,Fu:2019,Scherer:2019,Sachdev:2019}.  

\textit{Note added}
After submission of our manuscript, a related work \cite{hu:2019} appeared at the arXiv preprint server.
 
\begin{acknowledgments}
\textit{Acknowledgements}
We thank Risto Ojaj\"arvi for discussions.
This work was supported by the Academy of Finland under Projects No. 303351, No. 307419, No. 317118, No. 318987, and by the European Research Council (ERC-2013-AdG-340748-CODE).
L.L.  acknowledges the Aalto Centre for Quantum Engineering for support. 
A.J. acknowledges support from the Vilho, Yrj{\"o} and Kalle V{\"a}is{\"a}l{\"a} foundation.
Computing resources were provided by Triton cluster at Aalto University.
We acknowledge grants of computer capacity from the Finnish Grid and Cloud Infrastructure (persistent identifier urn:nbn:fi:research-infras-2016072533)
\end{acknowledgments}

\bibliography{bib_tbg}
\end{document}

\title{Superfluid weight and Berezinskii-Kosterlitz-Thouless transition temperature of twisted bilayer graphene: supplementary material}

\newcommand{\affiliationAalto}{Department of Applied Physics, Aalto University, P.O.Box 15100, 00076 Aalto, Finland}
\newcommand{\affiliationJYU}{Department of Physics and Nanoscience Center, University of Jyv\"askyl\"a, P.O. Box 35 (YFL), FI-40014 University of Jyv\"askyl\"a, Finland}
\newcommand{\affiliationTampere}{Computational Physics Laboratory, Physics Unit, Faculty of Engineering and Natural Sciences, Tampere University, P.O. Box 692, FI-33014 Tampere, Finland}

\author{A. Julku}
\affiliation{\affiliationAalto}
\author{T. J. Peltonen}
\affiliation{\affiliationJYU}
\author{L. Liang}
\affiliation{\affiliationAalto}
\affiliation{\affiliationTampere}
\author{T. T. Heikkil\"a}
\affiliation{\affiliationJYU}
\author{P. T\"orm\"a}
\affiliation{\affiliationAalto}

\date{\today}

%\begin{abstract}
%This is the SM.    
%\end{abstract}

\maketitle

\onecolumngrid

\section{Details of the Renormalized \moire (RM) and Dirac point (DP) models}
In this section we provide additional information on the details of our RM and DP models and how the order parameters are solved from both models.

\subsection{Renormalized \moire model (RM)}

\subsubsection{Computation of the order parameters}

Let us start by writing the Fermi-Hubbard model already presented in the main text:
\begin{align}
\label{fermihubbard}
H = \sum_{i\alpha j\beta\sigma}t_{i\alpha j\beta}c^\dag_{i\alpha\sigma}c_{j\beta\sigma} -\mu \sum_{i\alpha\sigma} c^\dag_{i\alpha\sigma}c_{i\alpha\sigma}  + H_{\textrm{int}},
\end{align}
where for the kinetic hopping amplitudes $t_{i\alpha j\beta}$ we use the parametrization provided by the Slater-Koster table of interatomic matrix elements~\cite{Slater:1954} for $p_z$ orbitals of the carbon atoms:
\begin{align}
\label{hopping}
t_{i\alpha j\beta} = t(\br) = t_0 \exp\Big[-\beta\frac{r - b}{b}\Big]\frac{x^2 + y^2}{r^2} +t_1 \exp\Big[ -\beta \frac{r-c_0}{b} \Big]\frac{z^2}{r^2},
\end{align}
where for simplicity we have denoted the distance between $\br_{i\alpha}$ and $\br_{j\beta}$ as $\br= [x,y,z]$. Here the $z$-axis is perpendicular with respect to the graphene layers. The first term in Eq.~\eqref{hopping} describes the intralayer hopping processes ($z=0$), whereas the interlayer processes are mainly described by the latter term. Here $b = a_0/\sqrt{3}$ is the distance between the nearest-neighbour carbon atoms, $a_0 = \SI{0.246}{nm}$ is the lattice constant of graphene, and $c_0 = \SI{0.335}{nm}$ is the interlayer distance. In our calculations we use parameters $t_0 = \SI{-2.7}{eV}$, $t_1 = \SI{0.297}{eV}$, and  $\beta = 7.2$. We restrict the interlayer hopping to the terms with $r< 4b$ and consider only the nearest-neighbour intralayer hopping terms.

For the interaction Hamiltonian, we use two different kinds of forms, namely the local attractive Hubbard interaction $H_{\textrm{int}} = J\sum_{i\alpha}c^\dag_{i\alpha\uparrow}c^\dag_{i\alpha\downarrow}c_{i\alpha\downarrow}c_{i\alpha\uparrow} \equiv H_{\textrm{loc}}$ and the resonance valence bond (RVB) type nearest-neighbour pairing potential $H_{\textrm{int}} = \frac{J}{2} \sum_{\langle i\alpha j\beta \rangle} h^\dag_{i\alpha j\beta} h_{i\alpha j\beta} \equiv H_{\textrm{RVB}}$, in order to see how the nature of the interaction Hamiltonian affects $D^s$ and $\TBKT$. Local interaction has been extensively used in the past graphene \cite{uchoa:2007,kopnin:2008,zhao:2006,hosseini:2015} and TBG studies \cite{peltonen:2018,wu:2018:2} to model $s$-wave superconductivity mediated by electron-phonon interaction.  Strictly speaking, in graphene the local (meaning also intrasublattice) interaction does not capture all implications of the attractive interaction mediated by electron--phonon coupling, but also the next-nearest neighbour interactions are often also included \cite{uchoa:2007,wu:2018:2}. In those works, it has been shown that the intersublattice coupling results to $d$-wave pairing. However, $s$-wave pairing was found to be the dominant pairing mediated by electron--phonon coupling, in which case the mean-field results of Ref. \onlinecite{wu:2018:2} are essentially the same as in the completely local model used in e.g. Ref. \onlinecite{peltonen:2018}.

In addition to local pairing potentials, also RVB and other non-local pairing schemes have been applied both in graphene \cite{baskaran:2002,black-schaffer:2007,pathak:2010,black-schaffer:2012,roy:2010,uchoa:2007,honerkamp:2008} and TBG studies \cite{su:2018}. To understand the RVB pairing scheme, let us note that it can be rewritten as $H_{\textrm{int}} = |J|\sum_{\langle i\alpha j\beta \rangle} \textbf{S}_{i\alpha} \vdot \textbf{S}_{j\beta} + 1/4 n_{i\alpha} n_{j\beta}$, where $\textbf{S}_{i\alpha}$ are spin operators, which is the usual Heisenberg antiferromagnetic Hamiltonian \cite{black-schaffer:2014}. For example, in the usual case of large repulsive on-site Coulomb interaction, the system Hamiltonian assumes the form of the $t-J$ model which has the double occupancy excluded and the interaction term is the aforementioned Heisenberg antiferromagnet. This leads to e.g. antiferromagnetic Mott-physics in case of a undoped simple square lattice as is well known \cite{lee:2006}. 

In our case we use RVB interaction but without the exclusion of the double occupancy, as in case of graphene the on-site repulsive Coulomb interaction between electrons is not necessarily large enough to justify the $t-J$ model. Early treatments of planar organic molecules of $\sigma\pi$-bonds were heavily based on the RVB approach \cite{pauling:book} and in 2002 it was suggested \cite{baskaran:2002} that the RVB approach would be a viable model to describe possible superconductivity in doped graphene and other similar organic layers. The argument for this is similar than in the case of the usual $t-J$ Hamiltonian: repulsive on-site Coulomb interaction results in two-body singlet correlations between neighbouring sites. However, as the on-site repulsion is not large enough, double occupancy is not entirely ruled out. Later in 2007 the RVB model was used in the mean-field level to show the possibility of $d+id$ superconductivity in doped graphene layers \cite{black-schaffer:2007}. In 2010, the authors of Ref. \onlinecite{pathak:2010}, inspired by the works of Refs. \onlinecite{baskaran:2002,black-schaffer:2007}, performed rigorous variational quantum Monte Carlo calculations by assuming local on-site Couloumb repulsion and showed that, indeed, doped graphene can support supercurrent with finite nearest-neighbour singlet pairing (RVB) correlations. Thus, non-local singlet pairing can be thought to originate from the Hubbard model of the graphene layers via the Coulomb electron-electron repulsion. Non-local singlet $d+id$ (and for some parameters also triplet $f$-wave) pairing in case of doped graphene was also obtained in Ref. \onlinecite{kiesel:2012} where functional renormalization group calculations were performed by assuming local and non-local repulsive Coulomb interactions. The authors of Ref. \onlinecite{kiesel:2012} showed that in graphene next-nearest-neighbour (NNN) and next-next-nearest-neighbour (NNNN) pairing schemes are possible. The nearest-neighbour RVB interaction used in our work is one of the simplest possible non-local pairing potentials but of course it would be interesting to also analyze these NNN and NNNN pairing schemes. This remains a topic of future studies.

To evaluate the chosen interaction Hamiltonians, we apply standard mean-field decoupling which yields
\begin{align}
&H_{\textrm{loc}} \approx \sum_{i\alpha} \Delta_{i\alpha} c_{i\alpha\uparrow}^\dag c_{i\alpha\downarrow}^\dag + \textrm{H.c.}, \label{interactions1} \\
&H_{\textrm{RVB}} \approx \sum_{\langle i\alpha j\beta \rangle} \Delta_{i\alpha j\beta} (c^\dag_{i\alpha\uparrow} c^\dag_{j\beta\downarrow} - c^\dag_{i\alpha\downarrow} c^\dag_{j\beta\uparrow}) + \textrm{H.c.} \label{interactions2},
\end{align}
where the order parameters are $\Delta_{i\alpha} = J \langle c_{i\alpha\downarrow} c_{i\alpha\uparrow}  \rangle$ and $\Delta_{i\alpha j\beta} = \frac{J}{2} \langle c_{j\beta\downarrow} c_{i\alpha\uparrow} - c_{j\beta\uparrow} c_{i\alpha\downarrow} \rangle$.

By using mean-field interaction terms \eqref{interactions1} and~\eqref{interactions2} and by exploiting the translational invariance, it is easy to rewrite the Hamiltonian \eqref{fermihubbard} in the momentum space as

\begin{align}
&H = \sum_{\bk} \Psi_{\bk} \mathcal{H}_{\bk} \Psi_{\bk}, \textrm{ where} \\
&\Psi_{\bk} = [c_{\bk\uparrow}, c_{-\bk\downarrow}]^T, \\
&c_{\bk\sigma} = [c_{\alpha=1,\bk,\sigma}, c_{\alpha=2,\bk,\sigma}, ..., c_{\alpha=M,\bk,\sigma}]^T,  \\
& c_{\alpha\bk\sigma} = \frac{1}{\sqrt{V}} \sum_{i\alpha} e^{\imag \bk \cdot \br_{i\alpha}} c_{i\alpha\sigma}
\textrm{, and} \\
&\mathcal{H}_{\bk} = \begin{bmatrix}
	\mathcal{H}_\uparrow(\bk) -\mu_\uparrow &  \Delta(\bk) \\
	\Delta^\dag(\bk) & -\mathcal{H}_\downarrow^*(-\bk) + \mu_\downarrow,
\end{bmatrix},
\end{align}
where the diagonal (off-diagonal) blocks are the Fourier transforms of hopping (pairing) terms. Here $M$ is the number of lattice sites per unit cell, $V$ is the total area of the system and $\bk$ belong to the unit cell of the reciprocal lattice.

By solving the BdG eigenproblem  $\mathcal{H}_{\bk} \ket{\psi_{i\bk}} = E_{i\bk} \ket{\psi_{i\bk}}$, we obtain the eigendecomposition  $\mathcal{H}_{\bk} =  V_{\bk} D_{\bk} V_{\bk}^\dag$. The diagonal matrix $D_{\bk}$ contains the eigenenergies $E_{i\bk}$, whereas the columns of the unitary matrix $V_{\bk}$ are the eigenstates $\ket{\psi_{i\bk}}$. One can then write down the self-consistent gap equations for the order parameters with the aid of $D_{\bk}$ and $V_{\bk}$. For the local interaction these read
\begin{align}
\label{eq:ge:loc}
\Delta_{i\alpha} = J \langle c_{i\alpha\downarrow} c_{i\alpha\uparrow}  \rangle =  \frac{J}{N} \sum_{\bk} \Big[ V_{\bk}  f(D_{\bk})V_{\bk}^\dag  \Big]_{\alpha, M+\beta}, 
\end{align} 
and correspondingly for the RVB
\begin{align}
\label{eq:ge:rvb}
\Delta_{i\alpha j\beta} = \frac{J}{2} \langle c_{j\beta\downarrow} c_{i\alpha\uparrow} - c_{j\beta\uparrow} c_{i\alpha\downarrow} \rangle = \frac{J}{2N} \sum_{\bk} \Big\{ e^{-\imag\bk \cdot \br^{\textrm{rel}}_{j\beta i\alpha}} \Big[ V_{\bk}  f(D_{\bk})V_{\bk}^\dag \Big]_{\alpha,M+\beta} - e^{\imag\bk \cdot \br^{\textrm{rel}}_{j\beta i\alpha}} \Big[ V_{\bk}  \big(\mathbb{1}_{2M}-f(D_{\bk})\big)V_{\bk}^\dag \Big]_{\beta,M+\alpha} \Big\},  
\end{align}
where $\br_{j\beta i\alpha}^{\textrm{rel}} \equiv \br_{j\beta} - \br_{i\alpha}$, $f$ is the Fermi-Dirac distribution, $N$ is the number of unit cells, and $\mathbb{1}_{2M}$ is a unity matrix of size $2M \times 2M$. The densities for each lattice sites can be solved from the following equations,
\begin{align}
\label{eq:densities}
& n_{i\alpha \uparrow} = \frac{1}{N} \sum_{\bk} \Big[ V_{\bk}  f(D_{\bk})V_{\bk}^\dag  \Big]_{\alpha, \alpha},  \\
& n_{i\alpha \downarrow} = \frac{1}{N} \sum_{\bk} \Big[ V_{\bk}  \big(\mathbb{1}_{2M}-f(D_{\bk})\big)V_{\bk}^\dag  \Big]_{M+\alpha, M+\alpha}.
\end{align}

The gap equations [Eqs.~\eqref{eq:ge:loc} or~\eqref{eq:ge:rvb}] and density equations~\eqref{eq:densities} are solved iteratively with the fixed-point iteration scheme by choosing a random initial ansatz for the order parameters. The iteration is terminated when   the order parameters and densities are converged to a stable solution of the gap equations. Due to the translational invariance, we can write $\Delta_{i\alpha} = \Delta_{\alpha}$, $\Delta_{i\alpha j\beta} = \Delta_{\alpha \beta}$ and $n_{i\alpha \sigma} = n_{\alpha \sigma}$. For local interaction there exist $M$ order parameters (for each lattice site in the unit cell) and in case of RVB there are $3M$ order parameters to be solved (for each nearest-neighbour bond).
From the obtained order parameters one can then compute the superfluid weight $D^s$ as explained in section~\ref{sec:ds}.

\subsubsection{Rescaling approximation}

When the bilayer graphene systems are twisted close to experimentally used magic angle  $\theta^* \approx 1^\circ$, the unit cell consists of around 12000--13000 lattice sites. Such a huge problem is computationally a rather heavy task and thus we decrease the number of lattice sites per unit cell, $M$, by applying a rescaling trick which keeps invariant two important observables, namely the Fermi velocity of a single graphene $v_F$ and the \moire superlattice periodicity $a$ while increasing the twist angle $\theta$ and thus decreasing $M$. More specifically, the Fermi velocity is proportional to $a_0$ and $t_0$, so that $v_F \propto a_0 t_0$. On the other hand, the \moire periodicity is $a = a_0/2 \sin(\theta/2)$. With this information, one can introduce the following rescaling under which $a$ and $v_F$ remain invariant \cite{su:2018,gonzalez-arraga:2017}:
\begin{align}
t_0'= \frac{t_0}{\lambda}, \quad a_0' = \lambda a_0, \quad c_0' = \lambda c_0, \quad \lambda = \frac{\sin \theta'/2}{\sin \theta/2},
\end{align}
where the primed quantites are the ones used in computations. With this trick one can apply much larger twist angles $\theta'$ than the usual magic angle $\theta \sim \SI{1}{\degree}$ and thus have much less lattice sites per \moire unit cell than at $\theta \sim 1^\circ$. The rescaling is characterized by the rescaling parameter $\lambda \geqslant 1$: larger $\lambda$ means more aggressive rescaling and larger $\theta'$, whereas $\lambda =1$ corresponds to the unscaled model.

Most importantly, the rescaling  is able to reproduce the flat bands and dispersive bands sufficiently well near the charge neutrality point as demonstrated in Fig.~\ref{fig:supp:1} where the low energy band structure is depicted for three different scaling parameters $\lambda$. The unscaled angle is chosen to be $\theta = \SI{1.0138}{\degree}$ and the rescaled angles are $\theta' = 4.4085^\circ$, $\theta' = \SI{1.8901}{\degree}$, and $\theta' = \SI{1.0178}{\degree}$, corresponding to the scaling factors of $\lambda = 4.3475$, $\lambda = 1.8643$, and $\lambda = 1.004$ (we do not use here $\lambda = 1$ as our twist $\theta = \SI{1.0138}{\degree}$ is not strictly commensurate, i.e. it strictly does not yield periodic structure, whereas our twist angles $\theta'$ used in the computations are always commensurate i.e. they strictly preserve the translational invariance.) We see that already a rather aggressive rescaling with $M=676$ and $\lambda = 4.3475$ is able to reproduce reasonably well the low energy band structure and a less aggressive rescaling of $\lambda = 1.8643$ is in practice identical to the unscaled band structure. From Fig.~\ref{fig:supp:1} we also see that rescaling amplifies the small electron-hole symmetry of the unscaled system. This explains why we see a fairly non-symmetric $\TBKT$ profile as a function of  filling for RM model as shown in Figs. 2(c)--(d) of the main text.

The rescaling approximation can be qualitatively understood by noting that under the rescaling the intraband hopping becomes smaller, i.e. the interband hopping terms become relatively more prominent and therefore one does not need to apply such a small twist angle to obtain flat band structures near the charge neutrality point.

\subsubsection{Choosing the twist angle}

More aggressive rescaling, i.e. larger $\lambda$, yields a smaller amount of lattice sites per \moire unit cell. However, $\lambda$ cannot be arbitrarily large as too strong rescaling cannot reproduce the original unscaled one-particle energy band structure. Feasibility of a specific rescaling $\lambda$ depends on the value of $\theta$. Some twist angles $\theta$ allow one to use more aggressive rescaling than some other twist angles.

\begin{figure}
\includegraphics[width=0.9\columnwidth]{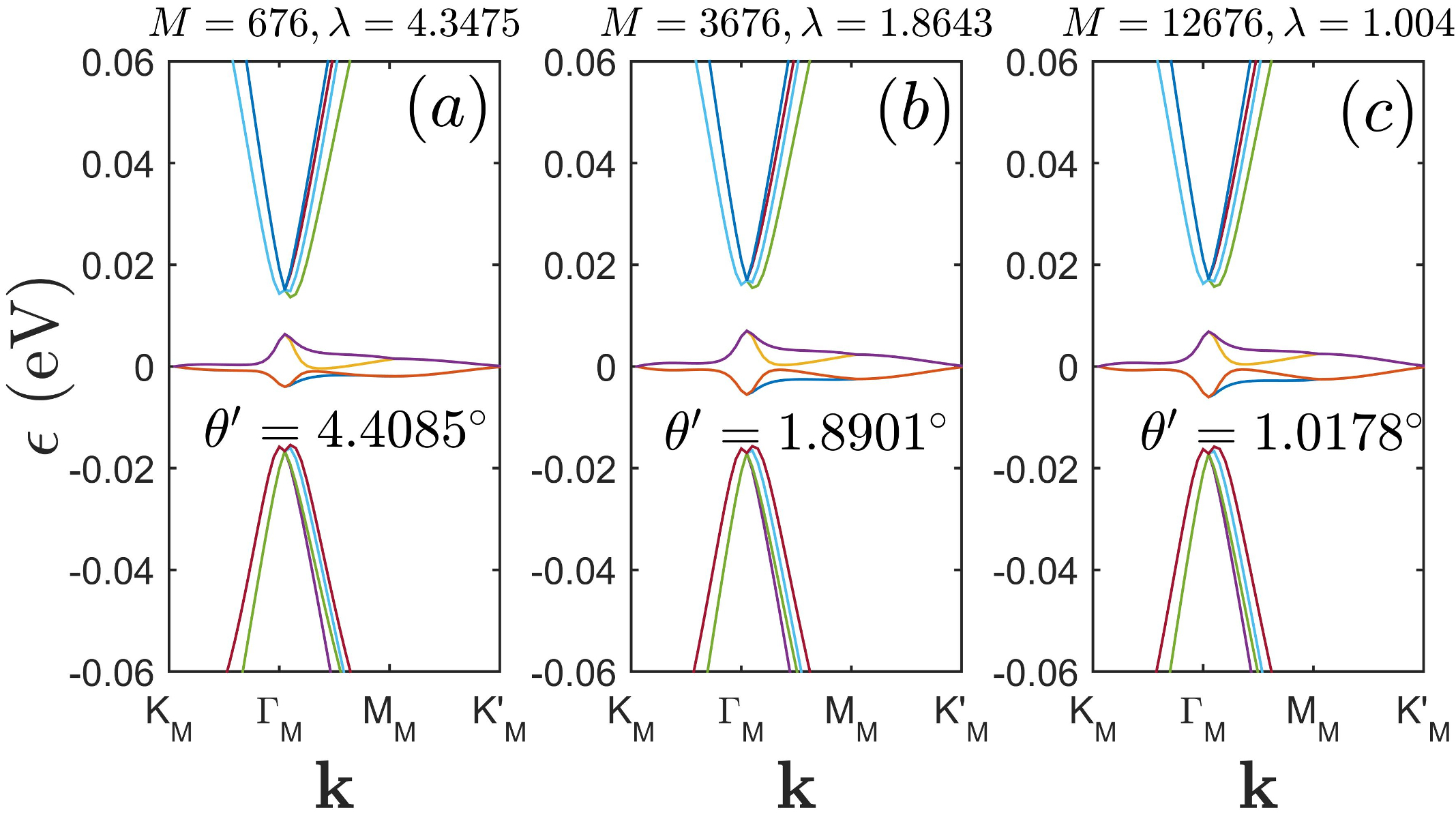}
\caption{\label{fig:supp:1}(a)--(c) Low energy band dispersions for $\theta = \SI{1.0138}{\degree}$ by using three different rescaling strengths. The most aggressive rescaling, namely $\lambda =4.3475$, slightly alters the shape of the flat bands but at the same time yields  a considerably easier problem to solve with only $676$ lattice sites per unit cell.}
\end{figure}

\begin{figure}
\includegraphics[width=1.0\columnwidth]{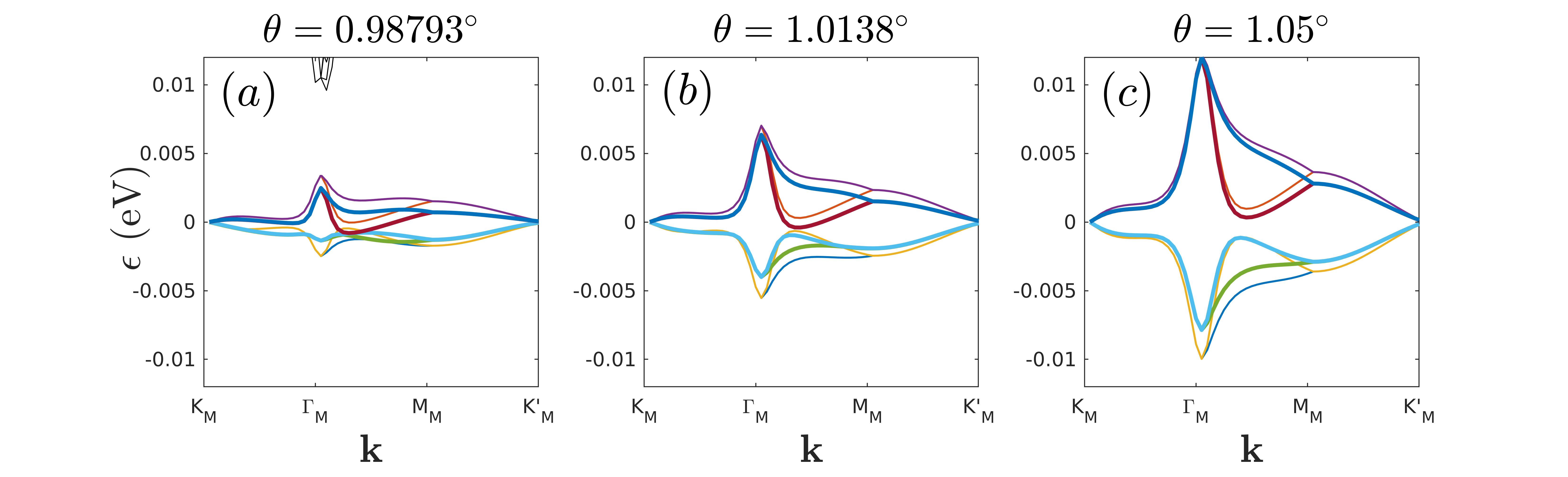}
\caption{\label{fig:supp:2}Flat band dispersion for three different twist angles computed by using two different rescaling strengths. Thick lines correspond to dispersions computed with strong rescaling that yields $M=676$, whereas thin lines correspond to rescaling of $M= 3676$.}
\end{figure}

The unscaled tight-binding model yields reasonably narrow bands near the charge neutrality point for the twist angles in the range of around $\theta \approx \SI{0.95}{\degree}\dots\SI{1.05}{\degree}$ so that the bandwidth of the flat bands at $\theta =\SI{1.05}{\degree}$ is around $\SI{20}{meV}$, whereas near $\SI{0.95}{\degree}$ it is less than $\SI{10}{meV}$. In this angle range also the band gaps between the dispersive and flat bands are notable. Based on these remarks, one is tempted to use angles near $\SI{0.95}{\degree}$ as there the bandwidth is at smallest. However, it turns out that to reproduce the shape of these extremely narrow flat bands near $\SI{0.95}{\degree}$ requires extremely mild rescaling and so one has to deal with a large number of lattice sites within a unit cell. On the contrary, for less narrow bands one can apply a more aggressive rescaling. In Fig.~\ref{fig:supp:2} we show the flat band dispersions for three different twist angles: $\theta =\SI{0.98793}{\degree}$ [Fig.~\ref{fig:supp:2}(a)], $\theta = \SI{1.0138}{\degree}$ [Fig.~\ref{fig:supp:2}(b)], and $\theta = \SI{1.05}{\degree}$ [Fig.~\ref{fig:supp:2}(c)]. The dispersions are plotted with two different rescaling angles: the thick lines correspond to the rescaling that yields $M= 676$ and the narrow lines to the rescaling with $M=3676$. The latter rescaling is sufficient to get fairly accurate band structures compared to the unscaled model. We see that more aggressive rescaling yields smaller bandwidths and at $\theta =\SI{0.98793}{\degree}$ alters the shape of the bands considerably. From Fig.~\ref{fig:supp:2}(a) one can see that with aggressive rescaling the third lowest flat band actually touches the two lowest bands. For larger $\theta$ the shapes of the flat bands remain rather invariant under the aggressive scaling. In these cases two lowest bands remain, apart at the Dirac points, isolated from the two upper flat bands and also the overall shape of the bands is fairly well reproduced. As argued in the next section, it is important to preserve the shapes of the energy bands essentially unaltered to obtain the isotropic SC state characterized by the mixed $(p+ip) + (d+id)$ pairing symmetry.

In our RM computations we use $\theta = \SI{1.0138}{\degree}$ and $\lambda \approx 4.3475$ which yields $M=676$ and $\theta' = \SI{4.4085}{\degree}$, i.e. we use the dispersions depicted in Fig.~\ref{fig:supp:1}(a) and~\ref{fig:supp:2}(b). This choice is a good compromise between the bandwidth and the shape of the rescaled bands. 

\begin{figure}
\includegraphics[width=1.0\columnwidth]{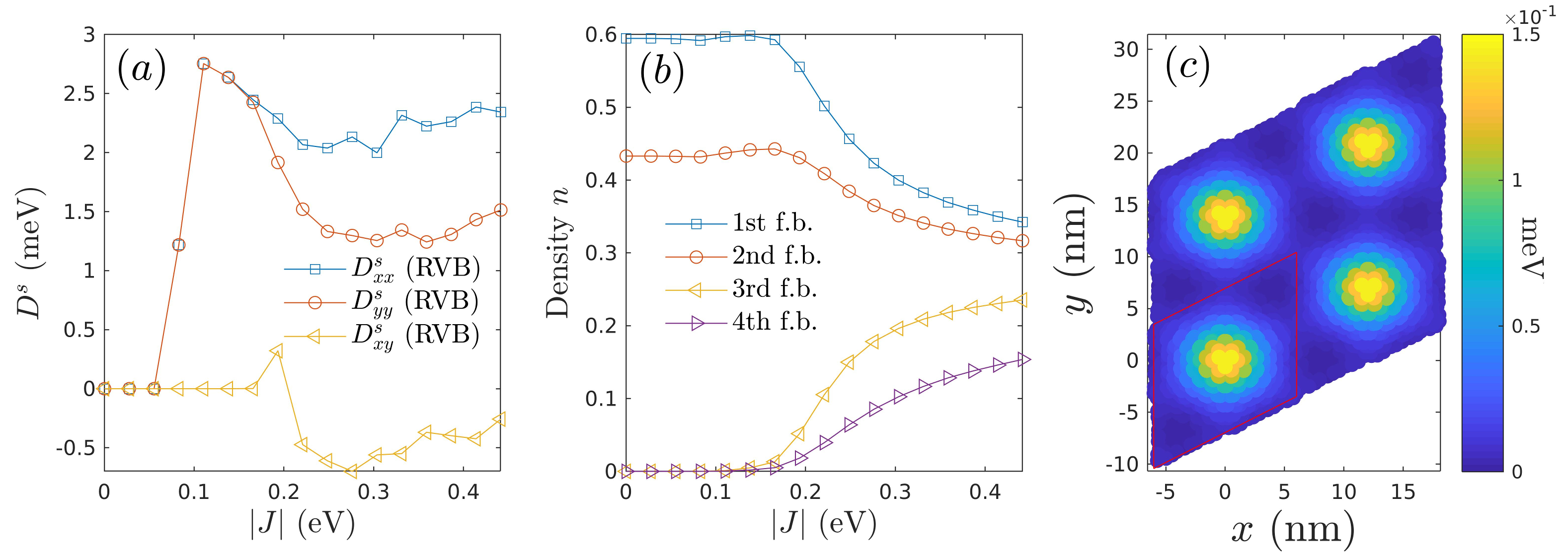}
\caption{\label{fig:supp:3}(a) Different spatial components of $D^s$ for RVB interaction at $\nu \approx -2$ as a function of the interaction strength $|J|$. (b) The corresponding densities of the flat bands. (c) The spatial profile of $\Delta_{i\alpha j\beta}$ at $|J| = \SI{0.11}{eV}$.}
\end{figure}

\subsubsection{Isotropic SC state with weak RVB interaction}

In the main text we showed that at the flat band regime the RVB interaction breaks the $C_3$-symmetry of the TBG lattice and consequently results in nematic SC states which manifest as an anisotropic superfluid weight. However, for weak enough RVB interactions, when the underlying symmetries of the lattice play a prominent role, one obtains isotropic SC states. This is shown in Fig.~\ref{fig:supp:3}(a) where we have reproduced the superfluid weight RVB results of Fig. 2(e) of the main text but this time (for clarity) as a function of the interaction strength $J$. One can see in the weak coupling regime the isotropic phase for which  $D^s_{xx} = D^s_{yy}$ and $D^s_{xy} = 0$. For some critical interaction strength the system then becomes nematic. 

This transition to the nematic phase is visible also in Fig.~\ref{fig:supp:3}(b) where we plot the densities of the four flat bands as a function of $|J|$. In the isotropic phase only the lower two flat bands are occupied whereas in the nematic phase the interaction is strong enough to redistribute some of the electrons to the upper flat bands. This is the reason why we are not using e.g. the twist angle $\theta = \SI{0.988}{\degree}$ depicted in Fig.~\ref{fig:supp:2}(a) for which the third lowest flat band actually touches the lowest flat band for a chosen rescaling strength. Due to this band touching, the electrons are redistributed to the upper flat bands with a vanishingly small interaction strength which prevents one to obtain the isotropic SC state. We emphasize that this band touching is an aberration caused by the rescaling approximation: for weaker rescaling (i.e. smaller $\lambda$) one should obtain the isotropic phase in the weak coupling regime also for $\theta = \SI{0.988}{\degree}$.

For completeness, the spatial profile of the order parameters of the isotropic phase is shown in Fig.~\ref{fig:supp:3}(c) for $|J|= \SI{0.11}{eV}$. In contrast to the isotropic phase resulting from the local interaction, the pairing symmetry here is not an $s$ wave but a mixed $(p+ip) + (d+id)$ wave as was shown in Ref.~\onlinecite{su:2018}.

\subsection{Dirac point model (DP)}
As described in Ref.~\onlinecite{peltonen:2018}, the $\vect{G},\vect{G}'$-component (vectors in the reciprocal superlattice) of the normal-state Hamiltonian matrix element $\mathcal{H}_{\rho\vect{k}}^0 \coloneqq \mathcal{H}_{\text{kin},\rho\vect{k}}^1 + \mathcal{H}_{\text{kin},\rho\vect{k}}^2 + \mathcal{H}_{\text{kin},\rho\vect{k}}^\perp - \mu\mathbb{1}$ at valley $\rho\in\{+,-\}$, $\vect{k}\in\MBZ$ is
\begin{equation}
    \mathcal{H}_{\rho\vect{k}}^0(\vect{G},\vect{G}') =
    \begin{pmatrix}
        [\hbar v_F \bm{\sigma}^\rho\vdot(\vect{k}+\vect{G}+\rho\Delta\vect{K}/2)-\mu]\delta_{\vect{G}\vect{G}'} & t_\perp^\rho(\vect{G}-\vect{G}') \\
        t_\perp^\rho(\vect{G}'-\vect{G})^\dagger & [\hbar v_F \bm{\sigma}_\theta^\rho\vdot(\vect{k}+\vect{G}-\rho\Delta\vect{K}/2)-\mu]\delta_{\vect{G}\vect{G}'}
    \end{pmatrix},
\end{equation}
where the matrix structure corresponds to the layer space, $\bm{\sigma}^\rho=(\rho\sigma_x,\sigma_y)$ consists of Pauli matrices acting in the sublattice space, $\bm{\sigma}_\theta^\rho = R(\theta)\bm{\sigma}^\rho$ is the $\theta$-rotated version of it, $v_F$ is the Fermi velocity of monolayer graphene, $\Delta\vect{K}=R(\theta)\vect{K}-\vect{K}$ is a vector from the graphene $\vect{K}$-point to its $\theta$-rotated counterpart, and $t_\perp^\rho(\vect{G})$ is a sublattice matrix containing a Fourier component of the interlayer coupling (times an exponential factor) with the elements
\begin{equation}
    t_\perp^{\rho,ss'}(\vect G) = \frac{1}{N} \sum_{\vect{r}\in\MUC} \e^{-\imag\vect{G}\cdot(\vect{r}+\delta_{sB}\bm{\delta}_1)} \e^{\imag\rho\vect{K}^\theta\cdot\bm{\delta}^{ss'}(\vect{r})} t_\perp(\bm{\delta}^{ss'}(\vect{r})).
\end{equation}
Here $\bm{\delta}^{ss'}(\vect{r})$ is the horizontal displacement vector between the site at $\vect{r}$, sublattice $s$ in layer 1 and the nearest-neighbor at sublattice $s'$ in layer 2. $\vect{\delta}_1$ denotes one of the nearest-neighbor vectors connecting the graphene A and B sublattices. The sum is over the graphene $A$ sublattice sites in the superlattice unit cell (the \moire unit cell MUC), and $N$ denotes the number of these sites. The interlayer coupling depends only on the (horizontal) distance $\bm{\delta}$ between the atoms, and is parametrized by a Slater-Koster parametrization as \cite{lopesdossantos:2007,lopesdossantos:2012}
\begin{equation}
    t_\perp(\delta) = \frac{1}{c_0^2+\delta^2} \left( c_0^2 V_{pp\sigma}\left(\sqrt{c_0^2+\delta^2}\right) + \delta^2 V_{pp\pi}\left(\sqrt{c_0^2+\delta^2}\right) \right),
\end{equation}
with
\begin{equation}
    V_{pp\sigma/\pi}(r) = \alpha_1^{\sigma/\pi} f_{pp\sigma/\pi}(r), \quad f_{pp\sigma/\pi}(r) = r^{-\alpha_2^{\sigma/\pi}} \exp(-\alpha_3^{\sigma/\pi}r^{\alpha_4^{\sigma/\pi}}).
\end{equation}
Here $c_0=\SI{3.35}{\angstrom}$ is the Bernal graphite interlayer distance, $a_0=\SI{2.461}{\angstrom}$ is the graphene lattice constant, and the $\alpha$ parameters are chosen as
\begin{equation*}
\begin{matrix*}[l]
    \alpha_1^\sigma = t_\perp^0/f_{pp\sigma}(c), &\alpha_2^\sigma = 0.7620, &\alpha_3^\sigma = 0.1624, &\alpha_4^\sigma = 2.3509, \\
    \alpha_1^\pi = t^0/f_{pp\pi}(a_0/\sqrt{3}), &\alpha_2^\pi = 1.2785, &\alpha_3^\pi = 0.1383, &\alpha_4^\pi = 3.4490,
\end{matrix*}
\end{equation*}
where $t^0=\SI{-3.08}{eV}$ is the intralayer nearest-neighbour hopping energy and $t_\perp^0=\SI{0.27}{eV}$ is the Bernal bilayer graphene nearest-neighbour hopping energy.

In the superconducting state we consider only the local interaction, in which case the $\vect{G},\vect{G}'$-component of the BdG Hamiltonian reads
\begin{equation}
    \mathcal{H}_{\rho\vect{k}}(\vect{G},\vect{G'}) =
    \begin{pmatrix}
        \mathcal{H}_{\rho\vect{k}}^0(\vect{G},\vect{G}') & \Delta(\vect{G}-\vect{G'}) \\
        \Delta^*(\vect{G}'-\vect{G}) & -\mathcal{H}_{\rho\vect{k}}^0(\vect{G},\vect{G}')
    \end{pmatrix},
\end{equation}
where the matrix structure corresponds to the Nambu space, and  the components of the superconducting order parameter $\Delta=\diag(\Delta_{1A},\Delta_{1B},\Delta_{2A},\Delta_{2B})$ are solved from the self-consistency equation
\begin{equation}
    \Delta_{ls}(\vect{G}) = \lambda \sum_{\rho,b,\vect{G}'} \int_\MBZ\frac{\dd{\vect{k}}}{(2\pi)^2} u_{\rho b\vect{k},ls}(\vect{G}') v_{\rho b\vect{k},ls}^*(\vect{G}'-\vect{G}) \tanh(\frac{E_{\rho b\vect{k}}}{2k_B T}).
\end{equation}
Here the band sum $b$ is calculated over the positive energy bands, $l\in\{1,2\}$ is the layer index, $s\in\{A,B\}$ is the sublattice index, and $\ket{\psi_{\rho b\vect{k}}(\vect{G})} = (\ket{u_{\rho b\vect{k}}(\vect{G})}, \ket{v_{\rho b\vect{k}}(\vect{G})})^\transpose$ [in Nambu space] with $\ket{u_{\rho b\bk}(\vect{G})} = (u_{\rho b\bk,1A}(\vect{G}),u_{\rho b\bk,1B}(\vect{G}),u_{\rho b\bk,2A}(\vect{G}),u_{\rho b\bk,2B}(\vect{G}))^\transpose$ and $E_{\rho b\vect{k}}$ are the eigenvectors and eigenenergies of the BdG equation
\begin{equation}
    \sum_{\vect{G}'} \mathcal{H}_{\rho\bk}(\vect{G},\vect{G}') \ket{\psi_{\rho b\bk}(\vect{G}')} = E_{\rho b\bk}\ket{\psi_{\rho b\bk}(\vect{G})} \quad\Leftrightarrow\quad \mathcal{H}_{\rho\bk} \ket{\psi_{\rho b\bk}} = E_{\rho b\bk} \ket{\psi_{\rho b\bk}}.
    \label{eq:BdG_equation_DP}
\end{equation}
The self-consistency equation is solved by the fixed point iteration scheme for a fixed chemical potential $\mu$.

For the BdG Hamiltonian we can calculate the total number density for a given chemical potential $\mu$ from
\begin{equation}
    n = 2\sum_{\rho,b} \int_{\MBZ}\frac{\dd{\bk}}{(2\pi)^2} \left[\braket{u_{\rho b\bk}}{u_{\rho b\bk}} f(E_{\rho b\bk}) + \braket{v_{\rho b\bk}}{v_{\rho b\bk}} (1-f(E_{\rho b\bk}))\right],
\end{equation}
where the factor of $2$ comes from spin, $f$  is the Fermi-Dirac distribution, and the band sum $b$ is calculated over the positive energy bands.

\section{Calculation of the superfluid weight}
\label{sec:ds}

In this section we go through very briefly the essential equations to compute $D^s$ in case of our RM and DP methods, show how $D^s$ can be split to conventional and geometric terms, and discuss why our results for $D^s$ are not in agreement with results of Ref.~\onlinecite{hazra:2018}.

We compute $D^s$ by using the linear response theory stated in Ref.~\onlinecite{liang:2017}. Therefore our starting point is the Fermi-Hubbard Hamiltonian of \eqref{fermihubbard}. To probe the system current response, we apply a spatially slowly varying vector potential $\vect{A}$ via the Peierls substitution such that the hopping amplitude $t_{i\alpha j\beta} \equiv t_{ab}$ ($a \equiv i\alpha$, $b\equiv j\beta$) is rewritten as $t_{ab}(\vect{A}) = t_{a b} e^{-i\int_{\br_{b}}^{\br_{a}} \vect{A}(\br) \cdot d\br} \approx t_{ab} e^{-i\vect{A}(\br_{ab}^{\textrm{CM}}) \cdot \br_{ab}^{\textrm{rel}}}$ (we set the Planck constant and the elementary charge to unity, i.e. $\hbar = e = 1$), where $\br_{a b}^{\textrm{CM}}= (\br_{a} + \br_{b})/2$ and $\br_{a b}^{\textrm{rel}}=\br_{a} - \br_{b}$. Then we expand the exponents up to second order so that our Hamiltonian becomes $H(\vect{A}) = H + \sum_\mu \sum_{a b} A_\mu (\br_{a b}^{\textrm{CM}}) j_\mu^p(a,b) + \frac{1}{2}\sum_{\mu\nu} \sum_{a b} A_\mu(\br_{a b}^{\textrm{CM}}) T_{\mu\nu}(a,b) A_\nu (\br_{a b}^{\textrm{CM}})$ . Here $j^p_\mu (a,b) = \sum_\sigma t_{ab} r^\textrm{rel}_{ab,\mu} c^\dag_{a\sigma} c_{b\sigma}$ is the paramagnetic current operator and $T_{\mu\nu} (a,b) A_\nu (\br_{a b}^{\textrm{CM}}) = \sum_\sigma t_{ab}   r^\textrm{rel}_{ab,\mu} r^\textrm{rel}_{ab,\nu} c^\dag_{a\sigma} c_{b\sigma} A_\nu(\br_{a b}^{\textrm{CM}})$ is the diamagnetic current operator. By using the expression for the total induced current density, $j_\mu(\br_{a b}^{\textrm{CM}}) = -\delta H(\vect{A})/\delta A_\mu(\br_{a b}^{\textrm{CM}})$, and linear response theory, we obtain in the momentum and frequency domain the relation $j_\mu(\bq,\omega) = -K_{\mu\nu}(\bq,\omega) A_\nu(\bq,\omega)$, where $K_{\mu\nu}$ is the current-current response function of the form
\begin{align}
\label{ccresponse}
K_{\mu\nu}(\bq,\omega) = \langle T_{\mu\nu} \rangle  - i\int_0^\infty dt e^{i\omega t} \langle [ j^p_\mu(\bq,t),j^p_\nu(-\bq,0)] \rangle,
\end{align}
and
\begin{align}
& T_{\mu\nu} = \sum_{\bk,\sigma} c^\dag_{\bk\sigma} \partial_\mu \partial_\nu \mathcal{H}_\sigma (\bk) c_{\bk\sigma}, \\
& j^p_{\mu}(\bq) = \sum_{\bk,\sigma} c^\dag_{\bk\sigma} \partial_\mu \mathcal{H}_\sigma (\bk + \bq/2) c_{\bk+\bq\sigma},
\end{align}
with $\partial_\mu \equiv \partial_{\bk_\mu}$, are the diamagnetic and paramagnetic current parts, respectively.

The superfluid weight $D^s$ is defined via the static Meissner effect ($\omega = 0$) and by taking the proper long wavelength limit of the transverse component of the current response function, see e.g. Refs.~\onlinecite{scalapino:1993,liang:2017}. In the mean-field level we can simply use the limit \cite{scalapino:1993}
\begin{align}
\label{ds}
D^s_{\mu\nu} = \lim_{\bq \rightarrow 0} \lim_{\omega  \rightarrow 0} K_{\mu\nu}(\bq,\omega).
\end{align}
This definition is equivalent with the one defined via the change of free energy due to the phase twist applied to the superconducting order parameter which leads to the form of $D^s_{\mu\nu} \propto \frac{\partial^2 \Omega(\vect{A})}{\partial A_\mu \partial A_\nu} \big|_{\vect{A}=0}$, where $\Omega$ is the grand canonical potential \cite{peotta:2015}.

Be deploying the mean-field theory and Green's function formalism, it was shown in Ref.~\onlinecite{liang:2017} for local Hubbard interactions that Eq.~\eqref{ds} leads to the following expression for $D^s$,
\begin{equation}
    D_{\mu\nu}^s = \frac{1}{V}\sum_{\bk,i,j} \frac{f(E_{j\bk})-f(E_{i\bk})}{E_{i\bk}-E_{j\bk}} (\mel{\psi_{i\bk}}{\partial_\mu\mathcal{H}_{\bk}}{\psi_{j\bk}} \mel{\psi_{j\bk}}{\partial_\nu\mathcal{H}_{\bk}}{\psi_{i\bk}} -\mel{\psi_{i\bk}}{\partial_\mu\mathcal{H}_{\bk}\tau_z}{\psi_{j\bk}} \mel{\psi_{j\bk}}{\tau_z\partial_\nu\mathcal{H}_{\bk}}{\psi_{i\bk}}),
    \label{ds2}
\end{equation}
where the eigenstates and eigenenergies are solved from the BdG equation $\mathcal{H}_{\bk}\ket{\psi_{i\bk}} = E_{i\bk}\ket{\psi_{i\bk}}$, $\tau_z$ is a Pauli matrix acting in Nambu space, $f$ is the Fermi-Dirac distribution, and $V$ is the area of the sample. The difference quotient is interpreted as $-f'(E_{i\bk})$ when $E_{i\bk}=E_{j\bk}$. In our TBG models the Hamiltonians are written in the superlattice-folded picture so that the $\bk$ sum is over the \moire Brillouin zone (MBZ) and the $i$ and $j$ sums are over the bands enumerating the eigenstates for each $\bk$. 

In case of the local interaction used in Ref.~\onlinecite{liang:2017}, the order parameters do not have momentum dependence and thus the derivatives $\partial_\mu \mathcal{H}_{\bk}$ are simply block diagonal matrices. However, for non-local interactions such as RVB used in our work, the order parameters depend on the momentum and thus the superfluid weight has a slightly different form, 
\begin{align}
    D_{\mu\nu}^s = \frac{1}{V}\sum_{\bk,i,j} \frac{f(E_{j\bk})-f(E_{i\bk})}{E_{i\bk}-E_{j\bk}} (&\mel{\psi_{i\bk}}{\partial_\mu\mathcal{H}_{\bk}(\Delta=0)}{\psi_{j\bk}} \mel{\psi_{j\bk}}{\partial_\nu\mathcal{H}_{\bk}}{\psi_{i\bk}} \notag\\
    -&\mel{\psi_{i\bk}}{\partial_\mu\mathcal{H}_{\bk}(\Delta=0)\tau_z}{\psi_{j\bk}} \mel{\psi_{j\bk}}{\tau_z\partial_\nu\mathcal{H}_{\bk}(\Delta=0)}{\psi_{i\bk}}).
    \label{ds3}
\end{align}
The only difference compared to Eq.~\eqref{ds2} is the derivatives of the order parameters in the diamagnetic part. However, in our case the order parameters are always really small compared to the kinetic terms and therefore we can in practice ignore extra terms arising from the derivatives of the order parameters. Therefore, in case of RM method we apply~\eqref{ds2} for both the local and RVB interaction schemes by taking $\partial_\mu \mathcal{H}_{\bk} = \partial_\mu \mathcal{H}_{\bk}(\Delta = 0)$.

In the DP model we assume that most of the contribution comes from states near the Dirac points, so that after writing everything in the valley-separated formalism, Eq.~\eqref{ds2} reads
\begin{equation}
    D_{\mu\nu}^s = \frac{1}{V}\sum_{\rho,\bk,i,j} \frac{f(E_{\rho j\bk})-f(E_{\rho i\bk})}{E_{\rho i\bk}-E_{\rho j\bk}} (\mel{\psi_{\rho i\bk}}{\partial_\mu\mathcal{H}_{\rho\bk}}{\psi_{\rho j\bk}} \mel{\psi_{\rho j\bk}}{\partial_\nu\mathcal{H}_{\rho\bk}}{\psi_{\rho i\bk}}
    -\mel{\psi_{\rho i\bk}}{\partial_\mu\mathcal{H}_{\rho\bk}\tau_z}{\psi_{\rho j\bk}} \mel{\psi_{\rho j\bk}}{\tau_z\partial_\nu\mathcal{H}_{\rho\bk}}{\psi_{\rho i\bk}}),
    \label{eq:superfluidweight_local_DP}
\end{equation}
where the eigenstates and eigenenergies are solved from the BdG equation \eqref{eq:BdG_equation_DP}. In the case of the DP model we use Eq.~\eqref{eq:superfluidweight_local_DP} to compute $D^s$.

\subsection{The form of superfluid weight in the presence of \texorpdfstring{$C_3$}{C3} rotational symmetry}
It can be shown that the superfluid weight is isotropic in the presence of $C_3$ rotational  symmetry.
 Let $\mathbf{e}_1$ be a unit vector, and $\mathbf{e}_2$ is obtained from $\mathbf{e}_1$ by a  $C_3$ rotation, i.e., $\mathbf{e}_2=R(\frac{2\pi}{3})\mathbf{e}_1$, and then $\mathbf{e}_3=R(\frac{2\pi}{3})\mathbf{e}_2=-\mathbf{e}_1-\mathbf{e}_2$. We write the superfluid weight tensor by using the coordinate vectors $\mathbf{e}_1$ and $\mathbf{e}_2$ as
\begin{eqnarray*}
D^s= D^s_{11}\mathbf{e}_1 \mathbf{e}_1+2D^s_{12}\mathbf{e}_1 \mathbf{e}_2+ D^s_{22}\mathbf{e}_2 \mathbf{e}_2.
\end{eqnarray*}
After a $C_3$ rotation, we get
\begin{eqnarray*}
C_3 D^s C^{-1}_3&=&D^s_{11}\mathbf{e}_2 \mathbf{e}_2+2D^s_{12}\mathbf{e}_2 (-\mathbf{e}_1-\mathbf{e}_2)+ D^s_{22}(-\mathbf{e}_1-\mathbf{e}_2 )(-\mathbf{e}_1-\mathbf{e}_2 ),\\
&=&D^s_{22}\mathbf{e}_1 \mathbf{e}_1+2(D^s_{22}-D^s_{12})\mathbf{e}_1 \mathbf{e}_2+ (D^s_{11}+D^s_{22}-2D^s_{12} )\mathbf{e}_2 \mathbf{e}_2.
\end{eqnarray*}
Since $C_3 D^s C^{-1}_3=D^s$, we find $D^s_{11}=D^s_{22}$ and $D^s_{12}=D^s_{11}/2$. 
In terms of  the Cartesian coordinates $\mathbf{e}_x$ and $\mathbf{e}_y$ ($\mathbf{e}_1=\mathbf{e}_x$ and $\mathbf{e}_2=-\frac{1}{2}\mathbf{e}_x+\frac{\sqrt{3}}{2}\mathbf{e}_y$), the superfluid weight becomes
\begin{eqnarray}
D^s&=&D^s_{11} (\mathbf{e}_1 \mathbf{e}_1 +\mathbf{e}_1 \mathbf{e}_2 +\mathbf{e}_2 \mathbf{e}_2), \notag\\
&=&D^s_{11} \left[ \mathbf{e}_x \mathbf{e}_x +\mathbf{e}_x \left(-\frac{\mathbf{e}_x}{2}+\frac{\sqrt{3}}{2}\mathbf{e}_y\right) + \left(-\frac{\mathbf{e}_x}{2}+\frac{\sqrt{3}}{2}\mathbf{e}_y\right) \left(-\frac{\mathbf{e}_x}{2}+\frac{\sqrt{3}}{2}\mathbf{e}_y\right) \right], \notag\\
&=&\frac{3D^s_{11}}{4}(\mathbf{e}_x \mathbf{e}_x +\mathbf{e}_y \mathbf{e}_y ),
\end{eqnarray}
which is isotropic.

\subsection{Geometric contribution and flat band superconductivity}

As TBG is an extremely complicated multiband system, it is highly instructive to decompose $D^s$ into the contributions of different one-particle Bloch states. We do this by using the method presented in Ref.~\onlinecite{liang:2017}, namely we expand the BdG states $\ket{\psi_{i\bk}}$ in the basis of Bloch functions by writing
\begin{align}
\ket{\psi_{i\bk}} = \sum_{m=1}^M \big(w_{+,im} \ket{+} \otimes \ket{m}_{\uparrow} + w_{-,im}\ket{-}\otimes \ket{m^*_-}_{\downarrow} \big),
\end{align}
where $\ket{m}_{\uparrow}$ [$\ket{m^*_-}_{\downarrow}$] is the eigenstate of $\mathcal{H}_\uparrow(\bk)$ [$\mathcal{H}^*_\downarrow(-\bk)$] with the eigenenergy $\epsilon_{\uparrow,m,\bk}$ [$\epsilon_{\downarrow,m,-\bk}$] and $\ket{\pm}$ are the eigenstates of $\tau_z$ with eigenvalues $\pm 1$. As in our case  we in practice always have $\partial_\mu \mathcal{H}_{\bk} = \partial_\mu \mathcal{H}_{\bk}(\Delta = 0)$, it is straightforward to rewrite $D^s$ of Eq.~\eqref{ds2} in the following form:
\begin{align}
\label{ds_decompose1}
D^s = 2 \sum_{\bk,i,j}\frac{f(E_{j\bk})-f(E_{i\bk})}{E_{i\bk}-E_{j\bk}} \Big[& \sum_{m_1,m_2} w^*_{+,im_1}w_{+,jm_2} {}_{\uparrow}\bra{m_1}\partial_\mu \mathcal{H}_{\uparrow}(\bk) \ket{m_2}_\uparrow \sum_{m_3,m_4} w^*_{-,jm_3}w_{-,im_4} {}_{\downarrow}\bra{m_{-3}^*}\partial_\nu \mathcal{H}_{\downarrow}^*(-\bk) \ket{m_{-4}^*}_\downarrow \nonumber \\
& + (\mu \leftrightarrow \nu) \Big].
\end{align}
We apply this expression when studying in Fig. 4 of the main text the superfluid weight by taking into account only the four flat bands or eight (4 flat, 4 dispersive) bands.

The matrix elements of the current operator can be further written as follows
\begin{align}
\label{ds_decompose2}
[j_{\mu,\sigma}(\bk)]_{mn} = {}_{\sigma} \bra{m} \partial_\mu \mathcal{H}_\sigma (\bk) \ket{n}_\sigma = \partial_\mu \epsilon_{\sigma,m,\bk} \delta_{mn} + (\epsilon_{\sigma,m,\bk} - \epsilon_{\sigma,n,\bk}) {}_\sigma \bra{\partial_\mu m} n\rangle_\sigma.   
\end{align} 
From Eqs.~\eqref{ds_decompose1} and \eqref{ds_decompose2} we see that there exist two different kinds of terms: the diagonal matrix elements of the current operator depend only on derivatives of the one-particle energy dispersions while the off-diagonal elements only on the momentum derivatives of the Bloch states. Thus $D^s$ can be split into two terms: the conventional part $D^s_{\textrm{conv}}$ that includes only the diagonal, i.e. intraband, current operator matrix elements ($m_1 = m_2$ and $m_3 = m_4$), and the geometric part $D^s_{\textrm{geom}}$ that includes off-diagonal, i.e. interband, current operator matrix elements so that $D^s = D^s_{\textrm{conv}} + D^s_{\textrm{geom}}$. The conventional part consists purely of the intraband current terms and is therefore zero for a single exactly flat band (as $\partial_\mu\epsilon_{\sigma,m,\bk} = 0$). Other way to see this is to note that $D^s_{\textrm{conv}}$ is inversely proportional to the effective mass of the electrons \cite{peotta:2015} which for an exactly flat band is infinite. Therefore non-zero superconductivity of a flat band is always a multiband property involving interband current processes between the flat band and other bands, i.e. finite $D^s_{\textrm{geom}}$. 

Roughly speaking, $D^s_{\textrm{conv}}$ scales with the bandwidth, whereas $D^s_{\textrm{geom}}$ scales with the interaction strength as larger interaction implies larger band mixing and thus more prominent interband processes. Therefore it is not surprising that we find a large geometric contribution in the flat band regime, as shown in the main text. Because of similar reasoning, it is understandable that interband terms between the flat and dispersive bands affecting $D^s_{\textrm{geom}}$ for stronger interactions are important, and that at the flat band regime dispersive bands cannot be discarded when computing the total superfluid weight.

The importance of $D^s_{\textrm{geom}}$ and the origin of the flat band superfluidity was for the first time addressed in Ref. \onlinecite{peotta:2015} where generic multiband Hubbard models were studied at the mean-field level in the presence of local Hubbard interaction (characterized by the coupling strength $J$) and time-reversal symmetry. The authors of Ref. \onlinecite{peotta:2015} considered an isolated flat band limit, i.e. the case where the Fermi surface lies within the flat band and interaction $|J|$  much smaller than the band gap $E_{\textrm{gap}}$ between the flat and other bands, i.e. $|J| \ll E_{\textrm{gap}}$. As the other bands are well separated from the flat band and the interaction is weak enough not to considerably mix the bands, the Cooper pairing in practice takes place only within the flat band. One is then tempted to perceive this limit as a single-band problem for which $D^s$ would be zero as $D^s_{\textrm{conv}}$ is zero for a flat band and $D^s_{\textrm{geom}}$ is zero for a single-band problem. However, from Eq. \eqref{ds_decompose2}, one can see that the geometric contribution actually scales as a function of $E_{\textrm{gap}}$ and one has to be careful when taking the isolated flat band limit. It was shown in Ref. \onlinecite{peotta:2015} that, indeed, the superconductivity of an isolated flat band is caused by the geometric superfluid weight term which, at low temperatures and with uniform local on-site pairing reads
\begin{align}
&D^s_{\textrm{geom},\mu\nu} \propto \Delta\int_{\textrm{B.Z.}}g_{\mu\nu}^{\textrm{f.b.}}(\bk) \dd{\bk}  \equiv \Delta\textrm{Re}[M_{\mu\nu}^{\textrm{f.b.}}], \textrm{ where}  \label{peottaresult} \\
&M_{\mu\nu}^{\textrm{f.b.}} \equiv \int_{\textrm{B.Z.}}B_{\mu\nu}^{\textrm{f.b.}}(\bk) \dd{\bk}. 
\end{align}
Here $g^{\textrm{f.b.}}(\bk) = \textrm{Re} \bra{\partial_\mu n_{\textrm{f.b.}}(\bk)}\big( 1- \ket{n_{\textrm{f.b.}}(\bk)} \bra{n_{\textrm{f.b.}}(\bk)} \big) \ket{\partial_\nu n_{\textrm{f.b.}}(\bk)}$ is the quantum metric of the flat band ($\ket{n_{\textrm{f.b.}}}$ are the Bloch states of the flat band, $\partial_\mu \equiv \partial/\partial k_\mu$) and $B^{\textrm{f.b.}}(\bk)$ is the corresponding quantum geometric tensor whose real (imaginary) part gives the quantum metric (Berry curvature) of the flat band \cite{berry:1989}. Similar results can be obtained also without TRS \cite{liang:2017}. Note that in \eqref{peottaresult} $D^s \propto \Delta$, i.e. the superfluid weight is linearly proportional to the pairing amplitude in the isolated flat band limit. This is similar to the behaviour of $D^s$ of TBG in the presence of local interaction as can be seen from Fig. 4 of the main text, according to which $D^s$ grows linearly when $\max |\Delta| \gtrsim \SI{2}{meV}$, implying that $D^s$ in this limit is dictated by the quantum metric. %This interpretation is reinforced by the fact that in this interaction regime the contribution of the dispersive bands to the Cooper pairing is small compared to the contribution of the narrow bands, similarly as in case of the aforementioned isolated flat band regime studied in Refs. \onlinecite{peotta:2015,liang:2017} where the pairing takes essentially place only within the flat band of interest.

An intriguing property of relation \eqref{peottaresult} is the fact that it can be evaluated with the Bloch states of the flat band only, even if the geometric contribution is a multiband process involving the interband matrix elements of the current operator. This is because the influence of the other bands arises implicitly from the form of the flat band Bloch states which are defined by the geometric properties of the quantum states of the whole lattice structure. It should be also emphasized that the existence of the other bands are required even though the Cooper pairing essentially takes place within the flat band and the pairing within the other bands is very small compared to the flat band. This can be reflected to TBG, where in the flat band regime of $\max |\Delta| \gtrsim \SI{2}{meV}$ the existence of the dispersive bands have to be taken into account even though the Cooper pairing occurs predominantly within the narrow bands only. Furthermore, even though the geometric contribution $D^s_{\textrm{geom}}$ consists of interband current terms, it does not require the interband pairing to be nonzero. Actually, in case of the Lieb lattice, the interband terms are in practice vanishingly small but $D^s_{\textrm{geom}}$ is large \cite{julku:2016}. The same applies to the TBG computations of this work: interband order parameters are in general negligible compared to the intraband order parameters but $D^s_{\textrm{geom}}$ can nevertheless be the dominant contribution.

To give an intuitive picture to Eq. \eqref{peottaresult} and the superfluidity of the flat bands in general, let us consider the quantum metric in a more general footing. To this end, let us introduce the infinitesimal distance between the quantum states of the $n$th energy band as follows
\begin{align}
\label{bures}
D^2_{n}(\bk,\bk+\dd{\bk}) = 1 - |\langle n(\bk) | n(\bk +\dd{\bk})\rangle|^2,    
\end{align}
which reaches its maximum value of unity when the states are not overlapping at all. One can define the quantum metric $g^n_{\mu\nu}(\bk)$ by expanding \eqref{bures} as
\begin{align}
D^2_{n,\mu\nu}(\bk,\bk+\dd{\bk}) = \sum_{\mu\nu}\frac{1}{2} g^n_{\mu\nu}(\bk) \dd{\bk_\mu} \dd{\bk_\nu}.
\end{align}
Here the higher order terms are ignored. Thus the quantum metric is related to the overlap of the quantum states. For example, in Ref. \cite{peotta:2015} it was shown that the superfluidity of a topologically non-trivial flat band is always positive as for a topological Bloch band one cannot construct the Wannier functions to be maximally localized with exponentially decaying amplitude \cite{peotta:2015,brouder:2007}. This implies finite delocalization and thus finite overlap between the Wannier functions. Therefore, superfluidity of, at least topological, flat bands can be explained by finite overlap of the Wannier functions which allows finite current transport. 

However, also topologically trivial flat bands can support supercurrent which is related to the finite quantum metric. To understand this, in Ref. \onlinecite{torma:2018} flat band superconductivity was approached from a different angle, namely via the two-body problem. Usually, in a non-flat band system, the Fermi sea is unstable towards the formation of bound pairs. In a flat band there does not exist a well-defined Fermi surface due to the degenerate states of the flat band. If this degeneracy is preserved in the presence of interactions, the existence of a bound state is not enough for superconductivity as then the condensation of Cooper pairs to a certain momentum state and the formation of a coherent superconducting state is not probable. In other words, in the presence of degenerate bound states, the effective mass of the Cooper pairs is infinite. The authors of Ref. \onlinecite{torma:2018} showed that, in the presence of interactions, the degeneracy can be lifted and the mass of the Cooper pairs can be finite when the Bloch states of the flat band have finite overlap and when their spatial derivatives are non-zero. Importantly, in case of the uniform pairing, the condition of overlapping quantum states reduces to the quantum metric results of Refs. \onlinecite{peotta:2015,liang:2017}. This connection relates the quantum metric directly to finite overlapping of the wavefunctions of the flat band. Moreover, the quantum metric is also present in a localization functional that describes the spread of the Wannier functions \cite{peotta:2015}. Finite quantum metric integral bounds the functional from below and thus implies finite spread and hence finite overlap between the Wannier functions.

\begin{figure}
\includegraphics[width=0.6\columnwidth]{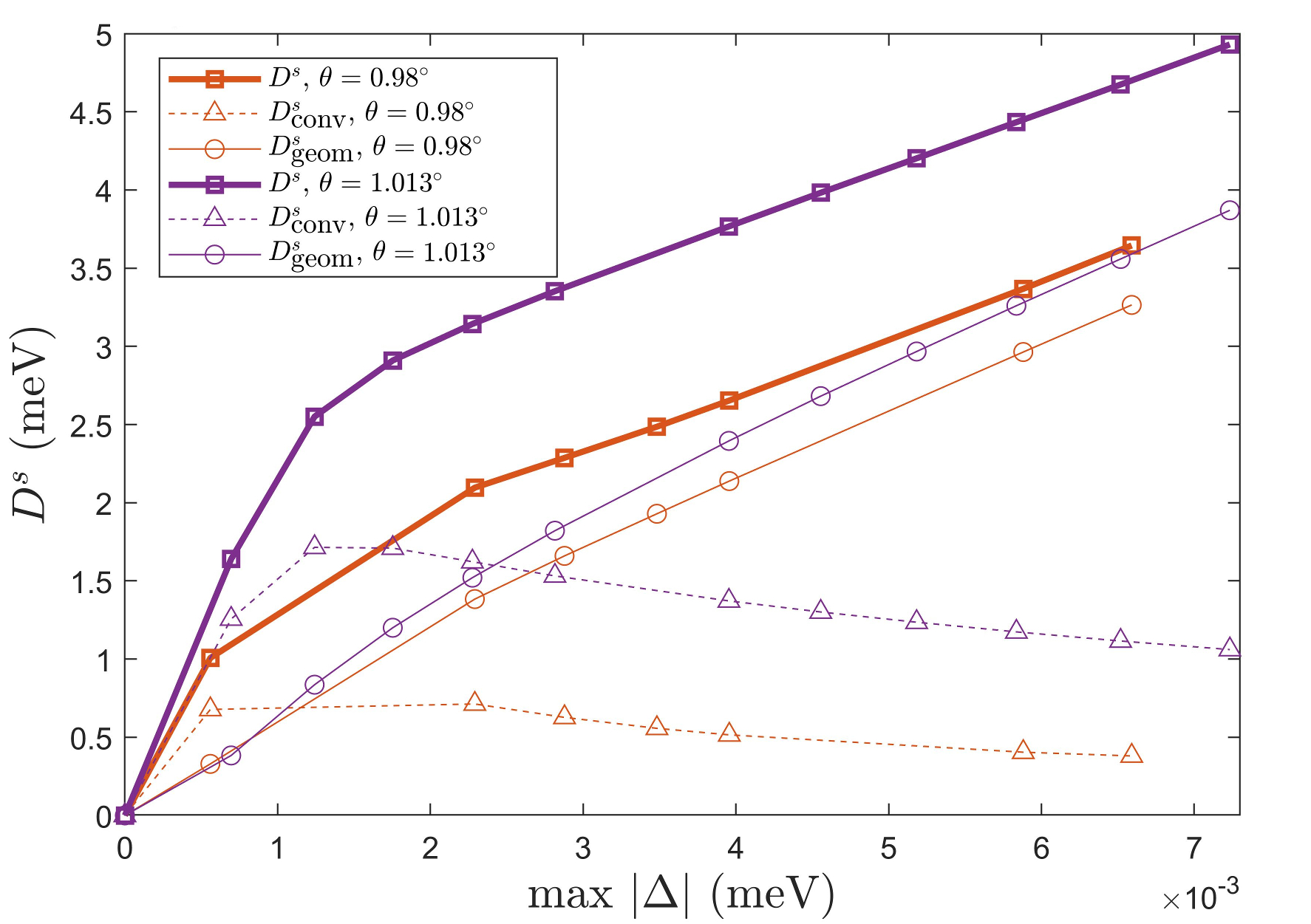}
\caption{$D^s$, $D^s_{\textrm{geom}}$, and $D^s_{\textrm{conv}}$ computed with RM method for two twist angles, $\theta = \SI{0.98}{\degree}$ (red lines) and $\theta = \SI{1.013}{\degree}$ (purple lines), at $\nu \approx -2$ and $T=1.5$ K as a function of the pairing strength in case of the local interaction. The results for $\theta = \SI{1.013}{\degree}$ are the same as those presented in Fig. 4(b) of the main text.}
\label{fig:supp:extra}
\end{figure}

In addition to the present work, the role of the geometric superfluid weight in case of TBG was also highlighted in Ref. \onlinecite{hu:2019} by using a TBG continuum model developed in Ref. \onlinecite{bistritzer:2010} where also other dispersive bands are taken into account. The main finding of Ref. \onlinecite{hu:2019} was that when the bandwidth of the narrow bands are minimized at the magic twist angle, the geometric contribution $D^s_{\textrm{geom}}$ is larger than the conventional term $D^s_{\textrm{conv}}$ for the interaction strength used in Ref. \onlinecite{hu:2019}. Correspondingly, by tuning the twist slightly off from the magic angle, the conventional term emerges as the main contribution, i.e. $D^s_{\textrm{conv}} > D^s_{\textrm{geom}}$. The conventional term was shown to depend heavily on the twist angle, whereas the geometric part was shown to be less sensitive to the twist. This is understandable as $D^s_{\textrm{conv}}$ depends strongly on the bandwidth of the narrow bands, decreasing as the bandwidth becomes smaller. Furthermore, as the interaction strength chosen in Ref. \onlinecite{hu:2019} was rather small, $\TBKT$ being around $0-\SI{2}{K}$, it is easy to comprehend that conventional term can dominate over the geometric one, as can be also seen e.g. from Figs. 4(a) and 4(b) of the main text where, at the regime of $\TBKT \sim 0-\SI{2}{K}$, $D^s_{\textrm{conv}}$ indeed is prominent. To see whether the RM method yields similar behavior, $D^s$, $D^s_{\textrm{geom}}$, and $D^s_{\textrm{conv}}$ computed with RM are plotted in Fig. \ref{fig:supp:extra} for two different twist angles, namely $\theta = \SI{1.013}{\degree}$ and $\theta = \SI{0.98}{\degree}$ as a function of the pairing strength at $\nu \approx -2$ in case of the local interaction. The bandwidth of the narrow band structure is slightly smaller for $\theta = \SI{0.98}{\degree}$ and, indeed, one can see from Fig. \ref{fig:supp:extra} that $D^s_{\textrm{conv}}$ is considerably smaller for $\theta = \SI{0.98}{\degree}$, whereas $D^s_{\textrm{geom}}$ is more or less the same for both angles. One can further see that there exists a weak-coupling regime where $D^s_{\textrm{geom}}> D^s_{\textrm{conv}}$ at $\theta = \SI{0.98}{\degree}$ but $D^s_{\textrm{geom}} < D^s_{\textrm{conv}}$ at $\theta = \SI{1.013}{\degree}$, reflecting the weak-coupling results of Ref. \onlinecite{hu:2019}.

In Ref. \onlinecite{xie:2019} a geometric lower bound for $D^s$ of TBG was derived in the absence of the dispersive bands and by assuming exactly flat bands. This lower bound was found to be proportional to the so-called Wilson loop winding number of the flat bands; a result that can be taken as yet another way to bound $D^s$ from below by the geometric properties of the quantum states, in addition to the bounds defined by the Chern number \cite{peotta:2015} and the Berry curvature \cite{liang:2017}. 
As the dispersive bands in Ref. \onlinecite{xie:2019} were ignored, the geometric contribution coming from the interband current terms between the flat and dispersive bands was not considered and the weak coupling limit was assumed. Thus the results of Ref. \onlinecite{xie:2019} cannot directly be related to the results presented in our work and in Ref. \onlinecite{hu:2019}, but all three works present strong arguments that geometric properties of the quantum states play a significant role in superconductivity of TBG. Particularly, our results highlight the necessity to consider the geometric  contribution, especially in case of stronger pairing interactions, when calculating the superfluid weight of TBG and other twisted multilayer systems. Therefore, being realizable in experiments, TBG can be potentially very important in terms of accessing and measuring the geometric contribution experimentally.

\subsection{Comparison to the ``upper'' limit of \texorpdfstring{$D^s$}{D} derived in Hazra \emph{et al.}}

In Ref.~\onlinecite{hazra:2018} the upper limit of $D^s$ and $\TBKT$ were computed for TBG system. The authors of Ref.~\onlinecite{hazra:2018} concluded giving an upper limit estimate of $D^s_{\textrm{max}} \sim \SI{1.5}{meV}$ (in our units, note that the superfluid weight definition of Ref.~\onlinecite{hazra:2018} differs from our definition by a factor of four), \emph{regardless} of the interaction mechanism or the interaction strength. Their estimate clearly contradicts with the results obtained by our two different models. The explanation for this disagreement is the use of oversimplified approximations in Ref.~\onlinecite{hazra:2018}. We go here briefly through their arguments for achieving the upper limit of $D^s$ and we argue why their upper limit for $D^s$ is not valid for arbitrary interaction strengths or mechanism.

The first important point is that the authors of Ref.~\onlinecite{hazra:2018} deploy an effective model, developed in Ref.~\onlinecite{koshino:2018}, that consists of only four flat bands. However, we showed in Fig. 4 of the main text that especially for strong interaction strengths the geometric contribution $D^s_{\textrm{geom}}$ arising from the off-diagonal matrix elements of the current operator [See Eq.~\eqref{ds_decompose2}] between the flat and dispersive bands is the most prominent part of the total superfluid weight. But the model used in Ref.~\onlinecite{hazra:2018} consists only of the flat bands, with dispersive bands being absent. As there are no dispersive bands implemented in their model, there cannot be any geometric contribution coming from the interband terms between flat and dispersive bands. Hence, the claim stating that the upper limit for $\TBKT$ derived in Ref.~\onlinecite{hazra:2018} holds for arbitrary interaction strength or interaction mechanism is readily shown to be invalid. This is not surprising: if the interaction strength is large enough, the dispersive bands become involved to the superconducting pairing, which is manifested by our results in Fig 4. of the main text.

To further highlight that the upper limit of Ref.~\onlinecite{hazra:2018} works only on the weak coupling regime, let us write down their argument. The starting point is the expressions \eqref{ccresponse}--\eqref{ds} which can be rewritten as
\begin{align}
D^s_{\mu\nu} = D^s_{\mu\nu, \textrm{dia}} + D^s_{\mu\nu, \textrm{para}},    
\end{align}
where $D^s_{\mu\nu, \textrm{dia}} = \langle T_{\mu\nu} \rangle$ is the diamagnetic part and correspondingly the paramagnetic contribution is  $D^s_{\mu\nu, \textrm{para}} = \lim_{\bq\rightarrow 0} \lim_{\omega\rightarrow 0} [- i\int_0^\infty dt e^{i\omega t} \langle [ j^p_\mu(\bq,t),j^p_\nu(-\bq,0)] \rangle]$. It can be shown that the paramagnetic part is always zero or negative, thus it follows that $D^s_{\mu\nu, \textrm{dia}} \geqslant D^s_{\mu\nu}$ (usually  in multiband systems the absolute values of dia- and paramagnetic parts are much larger than the absolute value of $D^s$). Therefore, the argument used by the authors of Ref.~\onlinecite{hazra:2018} is to compute the diamagnetic part $D^s_{\mu\nu, \textrm{dia}}$ to obtain the upper limit for the total superfluid weight $D^s$.

It is straightforward to rewrite the diamagnetic term in the following form,
\begin{align}
\label{eq:dia}
D^s_{\mu\nu,\textrm{dia}} = \langle T_{\mu\nu} \rangle = \sum_{m,m',\bk,\sigma} M^{-1}_{mm'}(\bk,\sigma) \langle c^\dag_{\bk\sigma m} c_{\bk \sigma m'} \rangle,    
\end{align}
where the inverse mass tensor is given by $M^{-1}_{mm'}(\bk,\sigma) = [U^\dag (\bk) \partial_\mu \partial_\nu \mathcal{H}_\sigma(\bk) U(\bk)]_{mm'}$. Here the columns of $U(\bk)$ are the one-particle Bloch states and $c_{\bk \sigma m'}$ is the annihilation operator for the Bloch state in the $m$th Bloch band of momentum $\bk$ and spin $\sigma$.

Now let us consider a situation where we are at the hole doping regime. The authors of Ref.~\onlinecite{hazra:2018} in this case assume that the two flat bands in the electron-doped side are empty. This is already an implicit assumption about the weak-coupling regime: for stronger interaction there exists finite electron occupation also in the upper flat bands, as can be seen in Fig.~\ref{fig:supp:3}(c) for example. Due to the Cauchy-Schwarz inequality $\langle c^\dag_{\bk\sigma m} c_{\bk \sigma m'} \rangle < \sqrt{ n_{\bk \sigma m} n_{\bk \sigma m'}}$, where $n_{\bk \sigma m} = \langle c^\dag_{\bk\sigma m} c_{\bk\sigma m} \rangle$, one can deduce then that  $\langle c^\dag_{\bk\sigma m} c_{\bk \sigma m'} \rangle =0$ if the band index $m$ or $m'$ refer to one of the two upper flat bands. There can still exist off-diagonal term if both $m$ and $m'$ refer to two hole-doping regime flat bands but also these off-diagonal elements are in Ref.~\onlinecite{hazra:2018} discarded. Thus the authors ignore the interband terms and end up having the form
$D^s_{\mu\nu\textrm{,dia}} = \sum_{m,\bk,\sigma} M^{-1}_{mm}(\bk,\sigma) \langle n_{\bk \sigma m} \rangle$. The occupation expectation value is then evaluated by assuming the step function $\langle n_{\bk \sigma m} \rangle = \Theta(\mu - \epsilon_m(\bk))$, i.e. by assuming the zero temperature and non-interacting limit for the occupation numbers. Therefore, their final upper bound for the superfluid weight of TBG system is $D^s_{\textrm{max}} = \sum_{m,\bk,\sigma} M^{-1}_{mm}(\bk,\sigma)  \Theta(\mu - \epsilon_m(\bk))$ and with this expression the authors obtain Fig.~1 shown in Ref.~\onlinecite{hazra:2018}, where the largest value for the superfluid weight (in our units) is around $\sim \SI{0.15}{meV}$. This is of the same order of magnitude than our weak-coupling results at very low temperatures, see for example Fig. 3(e) of the main text. This is not surprising as essentially all the assumptions done in Ref.~\onlinecite{hazra:2018} implicitly require weak interaction strengths. Thus rather than calling it the generic upper limit for $D^s$, the result of Ref.~\onlinecite{hazra:2018} can be taken as a reasonable estimate for $D^s$ in case of weak interactions. Note that this estimate is close to the superfluid weight of the RVB case shown in Fig. 4 of the main text but those results are evaluated at considerably higher temperatures of $T \approx \SI{1.5}{K}$.

\section{Superfluid weight and \texorpdfstring{$T_\mathrm{BKT}$}{TBKT} at the charge neutrality point}
In the main text we provided $\TBKT$ and $\TBKT/\textrm{max}|\Delta|$ as function of the pairing strength at half-filling of the hole-doping flat band regime, i.e. at $\nu \approx -2$. Here we provide, for completeness, similar plots for the case $\nu = 0$, i.e. at the charge neutrality point.
 
\begin{figure}
\includegraphics[width=1.0\columnwidth]{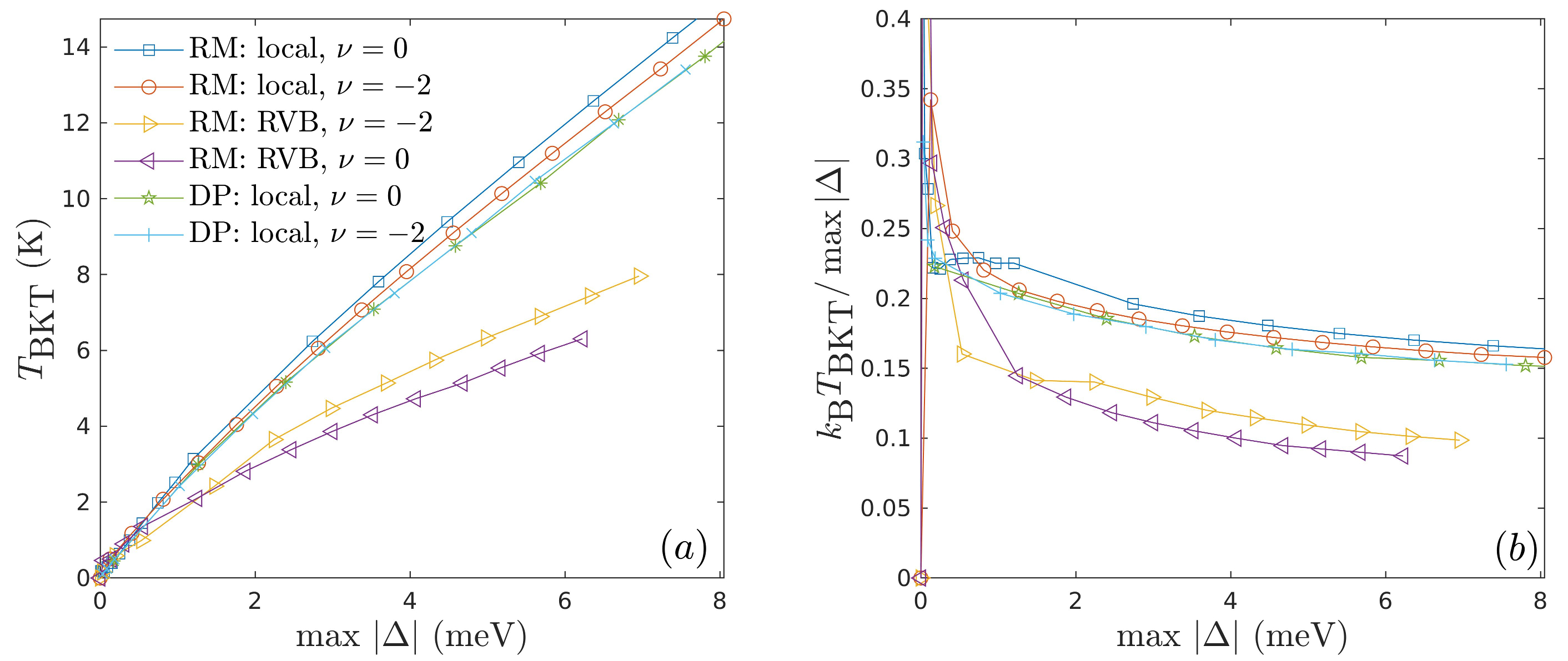}
\caption{\label{fig:supp:4}(a) $\TBKT$ as a function of max$|\Delta (T=0)|$ at $\nu=0$ and $\nu =-2$. (b) The corresponding results for $\TBKT/\textrm{max}|\Delta (T=0)|$.}
\end{figure}

In Fig.~\ref{fig:supp:4} we show,  as a function of max$|\Delta|$, $\TBKT$ [Fig.~\ref{fig:supp:4}(a)] and the ratio $\TBKT/\textrm{max}|\Delta|$ [Fig.~\ref{fig:supp:4}(b)] for $\nu=0$. For comparison, also the results of $\nu=-2$ are shown. One can see that both quantities behave very similarly for both fillings and especially in case of local interaction, $\TBKT/\textrm{max}|\Delta|$ seems to be rather independent on the filling in the flat band regime.   

\section{Superfluid weight as a function of the renormalization strength}

\begin{figure}
\includegraphics[width=1.0\columnwidth]{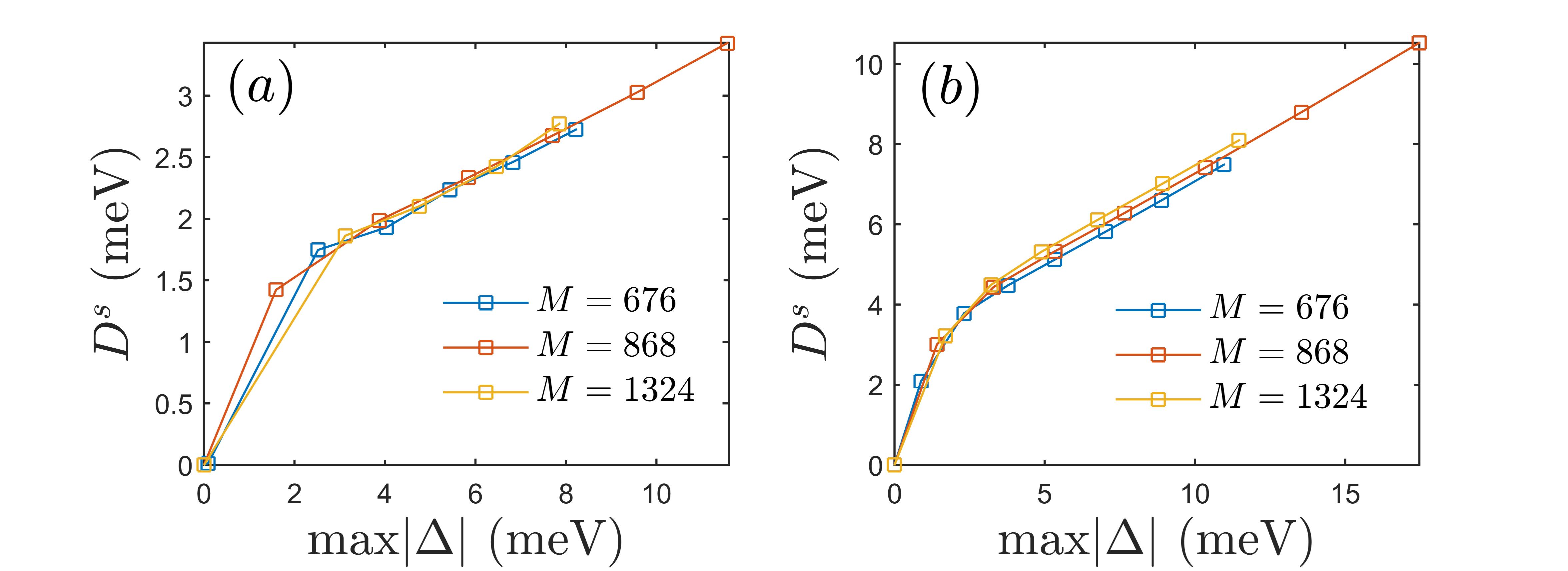}
\caption{\label{fig:supp:extra2}$D^s$ for three different renormalization strengths with $M=676$, $M=868$, and $M=1324$ in case of (a) RVB and (b) local interaction at $\nu \approx 0$ and $T=\SI{1.5}{K}$.}
\end{figure}

To demonstrate the validity of the renormalization scheme, in Fig. \ref{fig:supp:extra2} we plot $D^s$ for both local and RVB interaction schemes at $\nu \approx 0$ and $T=\SI{1.5}{K}$ for three different renormalization strengths with $M=676$, $M=868$, and $M=1324$, where $M$ is the number of lattice sites per unit cell. In all the calculations shown in the main text one has $M=676$. We see that the results remain more or less the same when $M$ is increased, i.e. when the strength of the renormalization is decreased.  Thus, using the renormalization of $M=676$ in the computations presented in the main text is justified. One should also note that the DP method (which has nothing to do with the renormalization method) yields similar results for $D^s$ than the RM method in case of the local interaction, as one can see from the results shown in the main text.

Heuristically, the renormalization method, yielding smaller amount of lattice sites within the \moire unit cell, can be thought as a coarse-grained model which has less degrees of freedom than the full microscopic model but which still features the same physics as the full model. The success of the renormalization method is not surprising as its main effect is to increase the relative strength of the interlayer coupling with respect to the intralayer coupling. Therefore, the rescaled model is still a twisted bilayer system which just has stronger interlayer coupling. Stronger interlayer hopping means that one can obtain the flat band dispersions with larger twist angles (and therefore with smaller $M$) than  with a system of smaller interlayer coupling. The idea is the same as in the experiment conducted by Yankowitz et al. \cite{yankowitz:2019} where the interlayer coupling was increased by hydrostatic pressure and thus the magic angle regime was reached for larger angles than in original experiments by Cao et al. \cite{cao:2018:1} where the pressure was not applied.

\bibliography{bib_tbg}